# Nernst effect of high-mobility Weyl electrons in NdAlSi enhanced by a Fermi surface nesting instability

Rinsuke Yamada[1,*], Takuya Nomoto[2], Atsushi Miyake[3], Toshihiro Terakawa[1], Akiko Kikkawa[4], Ryotaro Arita[2,4], Masashi Tokunaga[3], Yasujiro Taguchi[4], Yoshinori Tokura[1,4,5], Max Hirschberger[1,4,*]

[1]Department of Applied Physics, The University of Tokyo, Bunkyo-ku, Tokyo 113-8656, Japan

[2]Research Center for Advanced Science and Technology, The University of Tokyo, Komaba, Tokyo 153-8904, Japan

[3]The Institute for Solid State Physics, The University of Tokyo, Kashiwa 277-8581, Japan

[4]RIKEN Center for Emergent Matter Science (CEMS), Wako, Saitama 351-0198, Japan

[5]Tokyo College, The University of Tokyo, Bunkyo-ku, Tokyo 113-8656, Japan

*To whom correspondence should be addressed: ryamada@ap.t.u-tokyo.ac.jp or hirschberger@ap.t.u-tokyo.ac.jp

**Abstract**

The thermoelectric Nernst effect of solids converts heat flow to beneficial electronic voltages. Here, using a correlated topological semimetal with high carrier mobility $\mu$ in presence of magnetic fluctuations, we demonstrate an enhancement of the Nernst effect close to a magnetic phase transition. A magnetic instability in NdAlSi modifies the carrier relaxation time on 'hotspots' in momentum space, causing a strong band filling dependence of $\mu$. We quantitatively derive electronic band parameters from a novel two-band analysis of the Nernst effect $S_{xy}$, in good agreement with quantum oscillation measurements and band calculations. While the Nernst response of NdAlSi behaves much like conventional semimetals at high temperatures, an additional contribution $\Delta S_{xy}$ from electronic correlations appears just above the magnetic transition. Our work demonstrates **the engineering of the relaxation time, or the momentum-dependent self energy, to generate a large Nernst response independent of a material's carrier density, i.e. for metals, semimetals, and semiconductors with large $\mu$.**

# Main text

**Introduction**

Thermoelectric effects convert temperature gradients to useful electric power, and are broadly classified into two categories: First, the Seebeck effect, where a voltage is induced parallel to the heat gradient, and secondly the Nernst effect, where the voltage appears in a direction perpendicular to the heat flow. While both the Seebeck and Nernst effect of narrow-gap semiconductors have been intensively studied in the context of energy saving technology for more than two decades [1-3], the further technological potential of Nernst-type phenomena was pointed out recently [4]. The two most well-known contributions to this thermoelectric response are the normal Nernst effect from orbital motion of electrons, and the anomalous Nernst effect that appears in the ground state of magnetic materials, which is typically proportional to the magnetization [5-8]. Here, we highlight a magnetic enhancement of the former effect in a semimetal with high carrier mobility, a consequence of electronic correlations that has been rather little explored so far.

Thermoelectric phenomena are generally known to be very sensitive to details of the electronic structure in vicinity of the Fermi energy $E_F$, which separates occupied from unoccupied quantum states in a solid [9,10]. Specifically, the normal Nernst effect contains terms proportional to the filling-dependent change in carrier density, $dn/dE$, and terms proportional to the change in carrier relaxation (scattering) time, $d\tau/dE$ (see Methods). The $dn/dE$ term drives a large thermoelectric response only in materials with a low number of carriers, such as narrow-gap semiconductors, but the latter effect may largely enhance the thermoelectric response even in good metals [11-18]. One particularly well-established way to enhance band-filling dependence of the relaxation time, $d\tau/dE$, is by electron correlations, especially close to 'hot spots', or regions of enhanced scattering, on the electronic Fermi surface [11,19]. Meanwhile, a quantitative demonstration of enhanced Nernst signals from such hot spots has remained an open challenge, largely due to the low carrier mobility of many available correlated materials. To address this issue, we study the Nernst response of the topological semimetal NdAlSi with elevated carrier mobility [20-22] and hot spots of scattering at a Fermi surface nesting instability (Fig. 1). The results indicate that Nernst effects from $d\tau/dE$ become large and dominant in clean systems with highly mobile charge carriers coupled to collective modes, such as lattice waves or – our focus here – magnetic fluctuations.

Topological semimetals, where crystalline or time-reversal symmetries enforce degeneracies of electronic bulk bands and associated surface states in solids [17, 20-25], are suitable to demonstrate the proposed Nernst effect from $d\tau/dE$: They have large carrier mobilities $\mu$ due to protected band crossings, where the effective band mass approaches zero. Furthermore, their physical behaviour can be designed based on symmetry principles, and electron correlations can be introduced by chemical alloying while leaving crystal symmetries and related band crossings intact. Third and finally, transport properties can be studied quantitatively in such materials due to a low number of Fermi surface sheets with high carrier mobility, generating transport behaviour that is conveniently nonlinear as a function of magnetic field.

We choose the polar magnetic Weyl semimetals $R$AlSi/$R$AlGe ($R$ = rare earth) as promising materials for our proof-of-principle study, where high-mobility Weyl electrons coexist with complex magnetic order at low temperatures [20-22]. In these materials, Weyl fermions [23], that is singly degenerate linear band crossing points, are present due to breaking of inversion symmetry in the body-centered tetragonal space-group $I4_1md$ [Fig. 1(a)]. Contrary to inversion symmetric materials, where magnetic order may generate Weyl fermions and other band degeneracies [24], the electronic structure in polar and chiral topological semimetals forms a robust backdrop for the study of correlation phenomena, being changed but little at the magnetic transition point [20,21].

Here, we demonstrate that the thermoelectric response of a prototypical magnetic Weyl semimetal, NdAlSi, can be enhanced in vicinity of a magnetic instability. This correlation phenomenon benefits from the high carrier mobility of quasi-relativistic Weyl electrons, demonstrated here by a thorough analysis of the Nernst effect $S_{xy}$ in the high-temperature regime. Semiclassical transport theory, ignoring electron-electron scattering, also well reproduces $S_{xy}$ at the base temperature of our experiment. Meanwhile, an additional Nernst contribution $\Delta S_{xy}$ appears and is strongly enhanced just above the transition to long-range magnetic order at $T_N = 7$ K, where magnetic fluctuations are coupled to the Fermi surface of the electron gas. Through ab-initio calculations of the electronic structure, as well as careful evaluation of the experimental profile $\Delta S_{xy}(T)$ and its magnitude, we conclude that the additional term originates from the band-filling (energy) derivative of the carrier relaxation time, $d\tau/dE$, due to a magnetic nesting instability between high-mobility Weyl electrons. Finally, the results are placed in the wider context of thermoelectric materials with relaxation time contributions.

**Results**

**Electric transport and high mobility charge carriers**

The electronic band structure of NdAlSi is composed of several Fermi surfaces; Fig. 1(c) shows the result of our ab-initio band theory calculations, combined with the present electric transport experiments (see Supplementary Information). The Fermi surface includes two

Weyl-type sheets (β, Σ) with linear band dispersions, and two trivial pockets (γ, δ) with rather quadratic bands. Only the δ-pocket has electron-like transport characteristics; all others are hole-like (Figs. S2 and S3). In this work, we observe strong thermoelectric anomalies due to a Fermi surface nesting instability [26-28] towards a periodically modulated, helimagnetic state with ordering vector **Q** = (2/3, 2/3, 0) [Fig. 1(b), 22]. The ordering vector, shown as a fat yellow arrow in Fig. 1(d), causes strong magnetic scattering of high-mobility Weyl carriers in the Σ−pockets above the magnetic ordering temperature. Meanwhile, signatures of charge carriers in the other bands can also be clearly identified in our transport experiments.

We first construct band-theoretical models of the electronic structure in NdAlSi by electrical transport measurements. Here, electric current $J_x$, magnetic field $B_z$, and the transverse (Hall) electric field $E_y$ define a right-hand Cartesian frame along the *x*-, *z*-, and *y*-axes, respectively; this is shown in Fig. 2(a), inset. The Hall conductivity relates them according to $J_x = \sigma_{xy} E_y$. Field-dependent traces of $\sigma_{xy}$ show a dispersive shape with a sharp maximum around 2 Tesla, accompanied by pronounced Shubnikov-de Haas oscillations. Our two-carrier Drude fit, shown by black dashed lines in Fig. 2(a), well reproduces $\sigma_{xy}$ up to room temperature (see Methods). Figures 2(b) and 2(c) show electronic band parameters derived from the fit: $n_1$ and $n_2$ are positive, reaching values of $6.5 \times 10^{19}$ cm$^{-3}$ and $2.5 \times 10^{19}$ cm$^3$ at low temperature; this indicates coherent metallic conduction. Similar transport parameters are also obtained from the two-carrier fit of $\sigma_{xx}$ (see Supplementary Information). The carrier mobilities $\mu_1$ and $\mu_2$ rise monotonically upon cooling towards 2,000 cm$^2$/Vs and 6,000 cm$^2$/Vs at 8 K, high values among magnetic topological semimetals [29] that are testament to the low defect concentration in our single-crystalline samples.

It is helpful to consider the Hall effect in the context of Shubnikov-de Haas oscillations, which serve as a caliper for the Fermi surface cross-sectional area [30]. From the oscillation analysis, and consistent with Ref. [22], the carrier densities of the trivial γ Fermi surface pocket, and of β with linear Weyl-dispersion, are estimated as $6.6 \times 10^{19}$ cm$^{-3}$ and $1.0 \times 10^{19}$ cm$^3$, close to $n_1$ and $n_2$ from the Hall effect (see Supplementary Information). We conclude that the electrical Hall effect in NdAlSi is dominated by two hole-type Fermi surfaces, the trivial γ pocket with carrier density $n_1$, and the low-mass Weyl electrons of β with carrier density $n_2$ and exceedingly high carrier mobility $\mu_2$. Our main point of interest, the Weyl fermions in the Σ-pocket, may also represent a minority contribution to $n_2$ and $\mu_2$, but their fingerprints appear much more clearly in the thermoelectric response. All Fermi surface segments shown in Fig. 1(c) are clearly observed in quantum oscillation experiments that are well consistent with density functional theory calculations (Table S1).

**Thermoelectric phenomena: Seebeck effect**

Having established a basic understanding of electric transport in NdAlSi, we move on to the thermoelectric effects. Applying a temperature gradient ($-\partial_x T$) to the sample, voltages appear both along the *x*- and *y*-axes; these define the thermoelectric Seebeck and Nernst effects $S_{xx} = -E_x / |-\partial_x T|$ and $S_{xy} = E_y / |-\partial_x T|$. In Fig. 2(b), $S_{xx}$ is divided by temperature, a common way to correct for the reduced ability of electrons to carry entropy upon cooling. The Seebeck effect is positive – consistent with the sign of the Hall effect – and $S_{xx}/T$ is nearly independent of temperature above 50 K; a pronounced enhancement appears below 50 K, with maximum around $T_{max}$ = 15 K. Excluding the phonon-drag effect (Fig. S5), we proceed to interpret the thermoelectric phenomena in NdAlSi in terms of a thermally induced drift of charge carriers, using semiclassical transport theory. The Seebeck-enhancement at $T_{max}$ is closely related to the correlation-driven Nernst effect of Weyl electrons discussed in the following sections. Our focus here on the Hall and Nernst response of NdAlSi is motivated mostly by their enhanced sensitivity to high-mobility conduction electrons.

**Extracting band parameters from the Nernst effect**

It is well established that the thermoelectric effects $S_{xx}$ and $S_{xy}$ are much more sensitive to details of the electronic structure in close vicinity of the Fermi energy, as compared to electrical transport [9]. The Nernst effect is then a sensitive probe for the energy derivative of electronic band parameters in vicinity of the Fermi edge, such as $dn/dE$ for the carrier density and $d\tau/dE$ for the carrier relaxation time [10-17]. To separate the correlation-driven Nernst signal $\Delta S_{xy}$ originating from sizable $d\tau/dE$ in NdAlSi, we derive expressions for the thermoelectric response as a sum of various Fermi surface pockets. This kind of multi-band analysis of the thermoelectric effect has been attempted in the literature only rarely, e.g. for specific limiting cases in Ref. [31-34]. First, it is useful to normalize the Nernst effect by the electric resistivity $\rho_{xx}$; the experimental data at high and low temperatures are then well described by two terms, attributed to two Fermi surfaces γ and β

$$\frac{S_{xy}}{\rho_{xx}T} = A'_1 F(\mu_1, B) + A'_2 G(\mu_2, B), \qquad (1)$$

with prefactors $A_1$' and $A_2$', carrier mobilities $\mu_1$ and $\mu_2$, and dimensionless functions $F$ and $G$ for the γ- and β-pockets of the Fermi surface, respectively. Like the Hall conductivity, $F$, $G$ have pronounced maxima at intermediate magnetic fields if the carrier mobility is high. Our analysis shows that only at intermediate temperatures, an additional term $\Delta S_{xy}$ related to $d\tau/dE$ from correlations on the Σ pocket appears on top of this well-defined background; more details are given in Methods.

The background specified by Eq. (1), however, contains a lot of valuable information in itself, going far beyond what can be extracted from the Hall effect. Being highly nonlinear as a function of magnetic field, the Nernst signal in Fig. 3 gives the reasonable four fit parameters of Eq. (1) already at high temperature $T > 40$ K (see also Fig. S1). A good description of the data can be obtained even when fixing the ratios $A_1'/A_2' = 7.0$ and $\mu_2/\mu_1 = 3.1$ for the entire dataset, independent of temperature; the corresponding fit is shown by black dashed lines in Figs. 3(a) and 3(b). Encouragingly, the extracted carrier mobilities are close to the values in Fig. 2(c). The prefactors $A_1'$ and $A_2'$ meanwhile are directly related to $dn/dE$, that is to the effective band mass $m_\gamma$ of γ and Fermi velocity $v_\beta$ of β; these cannot be extracted from the Hall effect alone. Figures 3(i) and 3(j) show $m_\gamma$ and $v_\beta$ to be nearly independent of temperature, taking values of 0.25 $m_0$ ($m_0$ being the free electron mass) and $6.0 \times 10^5$ m/s, respectively, in quantitative agreement with our analysis of quantum oscillations at low temperatures (Figs. S2, S3; Table S1).

We further demonstrate that Eq. (1), without additional $\Delta S_{xy}$, well describes the experimental observations at the lowest temperatures, deep in the magnetically ordered state. The $\Delta S_{xy}$ term found in the temperature range 10 K to 30 K appears to be absent at the lowest temperatures. Extrapolating the band parameters $A_1'$, $A_2'$, $\mu_1$, $\mu_2$ as in Methods, we find an excellent, parameter-free description of the experimental data at $T = 2$ K, see Fig. 3(h). It is emphasized that, although the curve at the low temperature in Fig. 3(h) shows a sharp, step-like anomaly, this is not attributed to an anomalous Nernst effect from spin-orbit interactions [35], but it can be explained by conventional two carrier model. Instead, the semiclassical Lorenz force acting on drifting, high-mobility charge carriers in a magnetic field quantitatively explains the data. Significant deviations from Eq. (1) appear below 30 K, shown by red shaded areas in Figs. 3(c)-3(g), with the implication of an additional contribution $\Delta S_{xy}$ originating from relaxation time effects ($d\tau/dE$), strongly enhanced just above the Néel transition at $T_N = 7.4$ K.

**Nernst effect by scattering from magnetic fluctuations**

Why should the additional term $\Delta S_{xy}$ be ascribed to $d\tau/dE$, and thus to a magnetic instability driven by correlations of the electron gas in NdAlSi? We offer five pieces of evidence based on the profiles of $\Delta S_{xy}/\rho_{xx}T$ in Fig. 4, viewed in the context of ab-initio band theory calculations. First, the field-dependent profile $\Delta S_{xy}(B)$ is suppressed with increasing the field beyond 3-5 T. Such a signal cannot arise from the anomalous Nernst effect that is in proportion to the bulk magnetization [4-8]. Second, our analysis shows that $\Delta S_{xy}(B)$ is not likely due to additional enhancement of $A_1'$ or $A_2'$, since no anomaly appears in the carrier density and mobility around $T_N$ [Figs. 2(b) and 2(c)]. Third, the fact that $\Delta S_{xy}$ drops sharply at low temperature indicates that a potential change of the electronic band dispersion, e.g. due to zone-folding in the magnetically ordered state, cannot be its root cause. Also, this suppression suggests that a contribution from the anomalous Nernst effect due to Berry curvature of the topological band structure is unlikely [35]. Fourth, a scaling analysis of the Nernst effect as a function of magnetization in Fig. S4 rules out a competing scenario, where $\Delta S_{xy}$ is driven by the emergent magnetic field of spin fluctuations with left- or right-handed spin chirality [36-39]. Fifth and finally, we are well able to describe $\Delta S_{xy}$ in semiclassical theory by the expression

$$\frac{\Delta S_{xy}}{\rho_{xx} T} = B_3 \left( \frac{\pm 2\mu_3 B - \tan\theta_H (1 - (\mu_3 B)^2)}{(1 + (\mu_3 B)^2)^2} \right). \qquad (2)$$

where $\mu_3$, $\tan\theta_H = \rho_{yx}/\rho_{xx}$, and $B_3 \propto d\tau/dE$ are the carrier mobility, the (total) Hall angle, and a prefactor independent of the magnetic field, respectively (see Methods). Fits of $\Delta S_{xy}$ to Eq. (2) are shown as black dashed lines in Figs. 4(a)-4(c). On a quantitative level, we find $\mu_3/\mu_2 = 0.6$, consistent with the calculated ratio of Fermi velocities $v_\Sigma$ and $v_\beta$ of the two Weyl-type Fermi surfaces illustrated in Fig. 1(c).

Figure 4(e) shows the temperature-dependent profile of $d\tau/dE$ obtained from $B_3$, with a sharp maximum at $T_{max}$ = 15 K, implying characteristic broadening of the Fermi-Dirac distribution of $3.8k_B T_{max} \sim 3$ milli-electron Volt (meV), see Fig. S6. Figure 1(d) demonstrates consistency between this observation and the calculated band-filling dependence of the Weyl-type $\Sigma$ pocket, where the Fermi contour at $E_F$ = -3 meV just barely touches the yellow nesting plane, a hotspot of electron scattering introduced by the (fluctuating) magnetic order. Note that the sign $d\tau/dE > 0$ indicates enhanced of scattering of charge carriers below the Fermi energy.

**Discussion**

We argue above that the additional signal $\Delta S_{xy}$, appearing in a limited window of temperature, should be ascribed to charge carrier correlations and suppressed carrier relaxation time in proximity to a Fermi surface nesting instability. More quantitatively, the total relaxation rate of electrons is a sum of various terms according to Matthiessen's rule [9], with $\tau^{-1}(\mathbf{k}) = \tau_0^{-1} + \tau_{ph}^{-1} + \tau_{el-el}^{-1} + \tau_{el-m}^{-1}(\mathbf{k}) + \ldots$ where $\tau_0$, $\tau_{ph}$, and $\tau_{el-el}$ represent scattering from lattice defects, lattice vibrations, and electron-electron interaction respectively (see Supplementary Information). We assume for simplicity that only scattering of conduction electrons from magnetic fluctuations $\tau_{el-m}$ depends on the position $\mathbf{k}$ of an electronic state in momentum space. Then, the energy derivative that enters the formula for the Nernst effect is written as $d\tau/dE = d\tau/d\mathbf{k} \times d\mathbf{k}/dE$, and the latter factor is proportional to the inverse Fermi velocity $\mathbf{v}_F$. The derivative, and hence $\Delta S_{xy}$, become large when scattering from fluctuations constitutes a sizable portion of $\tau^{-1}(\mathbf{k})$, i.e. in clean materials with high carrier mobility. Figure 4(d) illustrates contour lines of the relaxation time $\tau(\mathbf{k})$ in momentum space, with a hot-spot of $\tau(\mathbf{k})$ that can be driven by magnetic fluctuations with propagation vector $\mathbf{Q} \approx (2/3, 2/3, 0)$ in NdAlSi.

Essential ingredients for large Nernst signals in the present mechanism are then (a) strong or at least moderately strong coupling between the electron gas and other (collective) degrees of freedom, impacting the carrier relaxation time and (b) the presence of conducting carriers with high mobility $\mu$. Although (b) is not exclusive to Weyl- or Dirac-electrons, or even to topological semimetals as a material class, the $\Sigma$ pocket in NdAlSi can well serve as a toy model system, where relaxation time effects enhance the thermoelectric Nernst voltage.

Cooling below $T_{\max} = 15$ K, all Weyl electrons of the $\Sigma$ pocket are now removed from the nesting hot-spot by more than the width of the Fermi-Dirac distribution. These electrons cannot participate in enhanced scattering from magnetic correlations – see Fig. 1(d) – and $\Delta S_{xy}$ begins to drop rapidly. The effect of $d\tau/dE$ on the thermoelectric Seebeck effect has been suggested in some recent work [11-18], while related phenomena in the Nernst effect remain largely unexplored. The existing literature is summarized in Table 1, which highlights the present Weyl semimetal NdAlSi's unique combination of high carrier mobility and sizable $d\tau/dE$. Sharp features in Nernst signal $S_{xy}(B)$ of NdAlSi thus reveal a small minority of high-mobility Weyl electrons, exposed to electron correlations by a Fermi surface nesting instability.

Recent insights into the materials engineering of topological matter have led to the identification of chiral and polar materials with robust linear band crossing points [21,25]. To these, correlations can be introduced via chemical composition tuning, while leaving intact the fundamental symmetries of the crystalline space group – and maintaining band degeneracies. The resulting high-mobility, correlated semimetals represent a family of weakly coupled materials that stand apart from the more established copper-oxide and iron-based superconductors [40-44], from layered chalcogenides with charge-density wave instabilities [45], or from heavy-fermion compounds [46]. For example, in the much more strongly correlated copper oxides, the interpretation of the thermoelectric Nernst effect [47-49] and the notion of hotspots of relaxation time on the Fermi surface [42] remain controversial. In contrast, NdAlSi and its relatives likely realize the weak-coupling scenario of Ruderman-Kittel-Kasuya-Yosida [26-28], where the momentum dependent self energy, or relaxation time $\tau(\mathbf{k})$, is calculated in first-order perturbation theory $\sim \int d^3 \mathbf{q}\, G(\mathbf{q})\, \chi(\mathbf{k}-\mathbf{q})$ from the electronic Green's function $G$ and the magnetic susceptibility $\chi(\mathbf{q})$. However, the present mechanism of an enhanced Nernst effect based on relaxation time engineering is generally applicable, providing a stimulus to investigate thermoelectric effects in as-yet less explored metallic materials, including clean systems with nesting instabilities such as elemental Cr or the $R$Te$_3$ layered CDW systems ($R$ = rare earth). In fact, we found an enhanced Nernst signal due to the relaxation time effect in another high-mobility semimetal, GdPtBi (see Supplementary Information), suggesting that enhancements of the thermoelectric response commonly appear in various correlated materials due to magnetic fluctuations close to the magnetic transition temperature. Even for the anomalous Nernst effect, previous work suggests a correlation of magnetic entropy (fluctuations) and the amplitude of the signal [50]. The present enhanced Nernst effect is particularly pronounced in crystals with high carrier mobility and low impurity concentration, criteria that are often fulfilled for large single-crystalline grains; yet even some polycrystalline materials can have very high carrier mobilities as well as large thermoelectric responses [51-53]. In addition, the relaxation time effect $d\tau/dE$ also has an impact not only on the Nernst effect but also on the Seebeck effect, which can be operated without the application of the magnetic field. In this way, thermoelectric phenomena generated via the relaxation time of charge carriers open a window onto the fundamental physics of electron correlation – but may also find real-world applications in energy harvesting using highly conductive metals, where conduction electrons are coupled to collective spin, orbital, or lattice fluctuations.

Add note: After submission of our manuscript, we became aware of a related work focused on quantum oscillations in $R$AlSi ($R$ = rare earth) [54].

## Methods

**Sample preparation and characterization**

Single-crystalline samples of NdAlSi were prepared by the Al flux method. Elemental Nd, Si, and Al elements were inserted into an $Al_2O_3$ crucible and sealed inside a quartz tube in vacuum. The starting composition (Nd:Si:Al) is 1:1:10. The tube was placed in a box furnace, heated from room temperature to 1000 ℃ in 5 hours, and kept there for 12 hours. Subsequently, it was slowly cooled to 700 ℃ in 50 hours (-6 ℃/h) and kept for 12 hours. The crucible was quickly removed from the furnace and centrifuged. Millimeter-sized, plate like crystals with the *c*-axis normal to the sample plane and the *a*-axis parallel to the plate's edges were obtained, which are characterized by sharp Laue x-ray diffraction patterns.

**Thermoelectric measurements**

Thermoelectric measurements on NdAlSi were performed in high vacuum (<1 × $10^{-4}$ Torr) using a Quantum Design Physical Property Measurement System (PPMS). The sample was fixed by Ag paste (Dupont) to a Cu block which serves as a heat sink. The Cu block is attached to the sample pack by using low-temperature adhesive (GE varnish). Thermal gradients were applied by applying current to a 1 kΩ chip resistor and temperature gradients on the rectangular-shaped sample (about 0.5 K/mm) were read by Nickel-Chromium/Constantan thermocouples. The thermocouples were attached to a thick gold wire by Stycast 2850FT epoxy and the thick Au wire in turn was attached to the side of the sample by silver paste. The thermoelectric voltage drop was measured through phosphor bronze wires (40 μm in diameter) attached to the side of the sample. The entire setup was covered by a copper tube to prevent radiative heating from outside.

**Ab-initio band calculations**

We perform spin density functional theory calculations in the ferromagnetic state of NdAlSi using the Vienna ab-initio simulation package [55]. We employ the exchange-correlation functional proposed by Perdew, Burke, and Ernzerhof [56], pseudopotentials with the projector augmented wave basis [57, 58], and use the experimental lattice parameters $a$ = 4.1972Å and $c$ = 14.4915Å [22]. The Hubbard $U$ correction with $U$ = 6.0 eV for Nd-4$f$ states and spin-orbit interactions are taken into account. A $16^3$ Monkhorst-Pack $k$-grid and 500 eV as a plane-wave cutoff are used for the self-consistent calculations. Then, we construct a tight-binding model using the Wannier90 package [59]. Here, 144 Bloch states are evaluated on a $8^3$ uniform **k**-grid. The trial orbitals are set to be Nd-$f$ and $d$, Al-$s$ and $p$, as well as Si-$s$ and $p$ orbitals. Fermi surface calculations are performed using a $240^3$ uniform **k**-grid based on the abovementioned tight-binding model.

**Mott relation and energy derivative of carrier relaxation time**

The electric and thermoelectric conductivity tensors define a material's current $J_i = \sigma_{ij} E_j$ and $J_i = \alpha_{ij}(-\partial_j T)$ in response to an applied electric field $E_j$ and to a temperature gradient $(-\partial_j T)$, respectively. Expressions for $\sigma_{ij}$, $\alpha_{ij}$ in a magnetic field are derived using semi-classical Boltzmann theory (9):

$$-\frac{\partial f_0}{\partial \varepsilon}\left[\left(\frac{\partial \zeta}{\partial \mathbf{x}}+\frac{\varepsilon-\zeta}{T}\frac{\partial T}{\partial \mathbf{x}}\right)\cdot \mathbf{v}-e\mathbf{E}\cdot \mathbf{v}\right]+\frac{\partial g_{\mathbf{k}}}{\partial \mathbf{k}}\cdot(\mathbf{v}\times \mathbf{B})=-\frac{g_{\mathbf{k}}}{\tau}. \qquad (3)$$

Here, $g_{\mathbf{k}}$ corresponds to a distortion of the equilibrium probability for occupying each quantum state, described by the Fermi-Dirac distribution function $f_0$. Further, $\varepsilon$, $\zeta$, $\mathbf{v}$, and $\tau$ are the electronic band dispersion, chemical potential, Fermi velocity, and carrier relaxation time, respectively. The magnetic field **B** is applied along **z** and the electric field **E** is parallel to **x**. We neglect any spatial dependence of $g_{\mathbf{k}}$, as well as Berry phase contributions. Integrating over all Fermi surface sheets, we derive the electric and thermoelectric currents as

$$\mathbf{J}_{\text{el}} = e^2 \sum_{bands}\int \frac{d^3\mathbf{k}}{(2\pi)^3}\left(\frac{\partial f_0}{\partial \varepsilon}\right)\left[\left(\frac{\mathbf{v}-\frac{e\tau}{m}\mathbf{v}\times \mathbf{B}}{1+\left(\frac{e\tau}{m}\right)^2 B^2}\right)\cdot \mathbf{E}\right]\tau_{\mathbf{k}}\mathbf{v} \qquad (4)$$

$$\mathbf{J}_{\text{th-el}} = e \sum_{bands}\int \frac{d^3\mathbf{k}}{(2\pi)^3}\left(\frac{\partial f_0}{\partial \varepsilon}\right)\left(\frac{\varepsilon-\zeta}{T}\right)\left[\left(\frac{\mathbf{v}-\frac{e\tau}{m}\mathbf{v}\times \mathbf{B}}{1+\left(\frac{e\tau}{m}\right)^2 B^2}\right)\cdot \frac{\partial T}{\partial \mathbf{x}}\right]\tau_{\mathbf{k}}\mathbf{v} \qquad (5)$$

The latter expression is further approximated by the crucial insight of Mott, who reduced the energy factor $(\varepsilon-\zeta)$ in Eq. (5) to an energy derivative over the other integrands [9]. The final expression depends both on the energy derivative of the carrier density, i.e. of the partial density of states (PDOS), as well as the energy derivative of $\tau_k$. The latter is the main focus of the present work. Note that the assumption of a unique value of $\tau_k$ for $\Sigma$ is a necessary but crude approximation, in case of the hotspot scenario discussed in the main text.

**Normal Nernst effect for more than one electronic band**

Consider the electric field **E** caused by the diffusion of charge carriers in a thermal gradient $\nabla T$ along **x** and a magnetic field **B** along **z**. The sign of the Nernst effect is then fixed by the convention that diffusion of superconducting vortices causes a positive $S_{xy}$, $S_{xy} = E_y/|\nabla T|$ [17,47]. Generally, the Nernst effect can have contributions from the orbital motion of carriers, from spin-orbit interactions in conjunction with the net magnetization **M** of the solid, and from higher order terms; for example from the spin chirality $S_i \cdot (S_j \times S_k)$. These are dubbed normal, anomalous, and topological Nernst effects, respectively. In the semi-classical regime and using the relaxation time approximation with isotropic $\tau$, the normal Nernst signal is expressed as,

$$S_{xy} = -\frac{\pi^2}{3}\frac{k_B^2}{e}T\rho_{xx}\sigma_{xx}\frac{\partial \tan\theta_H}{\partial \varepsilon}. \tag{6}$$

Here, $\tan\theta_H = \sigma_{xy}/\sigma_{xx}$ is the Hall angle. Further consider the Drude expressions for multi-band conduction, $\sigma_{xx} = \Sigma_i \, e|n_i|\mu_i/[1+(\mu_i B)^2]$ and $\sigma_{xy} = \Sigma_i \, e n_i \mu_i^2 B/[1+(\mu_i B)^2]$, with carrier densities $n_i$ and carrier mobilities $\mu_i = e\tau_i/m_i$; $e>0$ and $m_i$ are the fundamental charge and the effective band masses, respectively. By transposing $T$ and $\rho_{xx}$ to the left side, and transforming the remaining right hand side of Eq. (6), we derive

$$\frac{S_{xy}}{\rho_{xx}T} = \sum_i \left[ A_i \left(\frac{\pm\mu_i B - \tan\theta_H}{1+(\mu_i B)^2}\right) + B_i \left(\frac{\pm 2\mu_i B - \tan\theta_H(1-(\mu_i B)^2)}{(1+(\mu_i B)^2)^2}\right) \right]. \tag{7}$$

The upper and lower sign notation describes hole- and electron-type carriers, respectively. Here, $A_i = -\pi^2 k_B^2 \mu_i/3 \cdot dn_i/dE$ and $B_i = -\pi^2 k_B^2 n_i/3 \cdot d\mu_i/dE$ are the contributions from the energy derivatives of the carrier density and of the carrier mobility. Furthermore, by decomposing contributions from quadratic bands ($E_i^Q = \hbar^2 k_i^2/2m_i$) and linear bands ($E_i^L = \hbar k_i v_i$, while replacing $m_i \rightarrow E_{Fi}/v_{Fi}^2$ in the expression for the mobility), the following relation can be derived

$$\frac{S_{xy}}{\rho_{xx}T} = \sum_{i:\text{Quad}} \left[ A_i^Q \left(\frac{\pm\mu_i B - \tan\theta_H}{1+(\mu_i B)^2}\right) + B_i^Q \left(\frac{\pm 2\mu_i B - \tan\theta_H(1-(\mu_i B)^2)}{(1+(\mu_i B)^2)^2}\right) \right]$$

$$+ \sum_{i:\text{Line}} \left[ A_i^L \left(\frac{\pm\mu_i B - \tan\theta_H}{1+(\mu_i B)^2}\right) \right.$$

$$\left. + \left(\frac{A_i^L}{3} + B_i^L\right)\left(\frac{\pm 2\mu_i B - \tan\theta_H(1-(\mu_i B)^2)}{(1+(\mu_i B)^2)^2}\right) \right]. \tag{8}$$

Here, the coefficients for quadratic bands are $A_i^Q = \pm g_s g_d k_B^2 \mu_i m_i k_i / 6\hbar^2$ and $B_i^Q = -g_s g_d e k_B^2 k_i^3 / 18 m_i \cdot d\tau_i/dE$; the coefficients for linear bands are $A_i^L = \pm g_s g_d k_B^2 \mu_i k_i^2 / 6\hbar v_i$, $B_i^L = -g_s g_d e k_B^2 v_i k_i^2 / 18\hbar \cdot d\tau_i/dE$, where $g_s$ and $g_d$ are spin and orbital degeneracy of each Fermi surface pocket, respectively.

In the specific case of NdAlSi, the Nernst effect at high and low temperatures is well described by parallel conduction of a trivial hole pocket $\gamma$ ($E_\gamma = \hbar^2 k_\gamma^2 / 2m_\gamma$) and a Weyl-type hole pocket $\beta$ ($E_\beta = \hbar k_\beta v_\beta$). At intermediate temperatures, additional contributions to $S_{xy}$ that are not from the energy-dependent carrier density are summarized in the term $\Delta S_{xy}/\rho_{xx}T$,

$$\frac{S_{xy}}{\rho_{xx}T} = A_\gamma \left( \frac{\pm \mu_\gamma B - \tan\theta_H}{1 + (\mu_\gamma B)^2} \right) + A_\beta \left( \frac{\pm \mu_\beta B \left(5 + 3(\mu_\beta B)^2\right) - 2\tan\theta_H \left(2 + 1(\mu_\beta B)^2\right)}{3\left(1 + (\mu_\beta B)^2\right)^2} \right)$$

$$+ \frac{\Delta S_{xy}}{\rho_{xx}T}, \tag{9}$$

where the coefficients are $A_\gamma = \pm g_s g_d k_B^2 \mu_\gamma m_\gamma k_\gamma / 6\hbar^2$ and $A_\beta = \pm g_s g_d k_B^2 \mu_\beta k_\beta^2 / 6\hbar v_\beta$. Here, we define $A_\gamma' = A_\gamma/\mu_\gamma$ and $A_\beta' = A_\beta/\mu_\beta$. We set the spin and orbital degeneracies of $\gamma$ ($\beta$) to $g_s = 2$ ($g_s = 1$) and $g_d = 4$ ($g_d = 4$), respectively, for the analysis at high temperatures; this is consistent with the large spin-orbit splitting of $\beta$ in the nonmagnetic state of polar NdAlSi.

We obtain an excellent description of $S_{xy}$ by Eq. (9) with four free parameters ($A_\gamma'$, $A_\beta'$, $\mu_\gamma$, and $\mu_\beta$) at high temperatures above 40 K. However, features in $S_{xy}$ become less sharp at $T > 70$ K, so that a reduction in the number of free parameters was found to be helpful to stabilize the fit. We fixed the ratios $A_\gamma'/A_\beta' \sim 7.0$ and $\mu_\beta/\mu_\gamma \sim 3.1$ independent of $T$, retaining a very good description of the experimental data. In particular, all parameters extracted from the semiclassical model of $S_{xy}$ have distinct physical meaning and can be cross-referenced with Hall effect and quantum oscillation experiments.

We extend the analysis according to Eq. (9) to $T < 40$ K by extrapolating the band parameters $A_\gamma'$, $A_\beta'$, $\mu_\gamma$, and $\mu_\beta$ as follows. Firstly, $A_\beta' = \pm g_s g_d k_B^2 \hbar k_\beta^2 / v_\beta$ is set to be 0.45 independent of temperature, as changes of the electronic structure are moderate in this regime [see Figs. 2(b) and 3(j)]. Also, the carrier mobilities are extrapolated to low $T$, being proportional to the longitudinal conductivity $\sigma_{xx}$. Furthermore, we assume that the spin degeneracy ($g_s$) of the $\gamma$ pocket smoothly changes from unity at 2 K to $g_s = 2$ at 40 K, considering that the exchange splitting of electronic bands is enhanced at low temperatures. Meanwhile, $g_s$ of $\beta$ is set to be unity at all temperatures, reflecting its small Fermi surface volume and large spin-orbit splitting even in the paramagnetic state. In this way, we estimated the first and second term in Eq. (9), and isolated $\Delta S_{xy}$ at intermediate temperatures.

**Two carrier fit of the Hall conductivity**

We model the Hall conductivity $\sigma_{xy}$ with a two-carrier Drude expression as [9]

$$\sigma_{xy} = n_1 e \mu_1 \frac{\mu_1 B}{1 + (\mu_1 B)^2} + n_2 e \mu_2 \frac{\mu_2 B}{1 + (\mu_2 B)^2}. \quad (10)$$

Here, $n_1$ and $n_2$ ($\mu_1$ and $\mu_2$) are the carrier densities (carrier mobilities). The resulting carrier densities $n_1$ and $n_2$ show good agreement with quantum oscillations of the $\gamma$ and $\beta$ pockets, as discussed in the main text.

# Data availability

The data supporting the findings of this study are available from the corresponding authors upon reasonable request.

## Acknowledgements

We acknowledge H. Isobe, K. Ueda, N. Nagaosa, and T.-h. Arima for fruitful discussion. This work was supported by JSPS KAKENHI Grant Nos. JP21K13877, JP22H04463, JP22K20348, JP23H05431, and 23K13057 as well as JST CREST Grant Number JPMJCR1874 and JPMJCR20T1 (Japan) and JST FOREST JPMJFR2238 (Japan) and JST-PREST JPMJPR20L7. The authors are grateful for support by the Fujimori Science and Technology Foundation, the New Materials and Information Fofaundation, the Murata Science Foundation, the Mizuho Foundation for the

Promotion of Sciences, the Yamada Science Foundation, the Hattori Hokokai Foundation, the Iketani Science and Technology Foundation, the Mazda Foundation, the Casio Science Promotion Foundation, the Takayanagi Foundation, the Foundation for Promotion of Material Science and Technology of Japan (MST Foundation), and Yashima Environment Technology Foundation.

## Author contributions

R.A., Y.Ta, Y.To., and M.H. conceived and supervised the project. R.Y. and A.K. synthesized and characterized the single-crystals. R.Y., T.T. and M.H. carried out and analyzed the electrical and thermoelectric transport measurements. R.Y., A.M. and M.T. conducted tunnel diode oscillator (TDO) measurements under high magnetic fields. T.N. and R.A. performed ab-initio calculations. R.Y. and M.H. wrote the manuscript, with the support of T.N. and R.A. All authors discussed the results and commented on the manuscript.

## Competing interests

The authors declare no competing interests.

# Main Text Figures

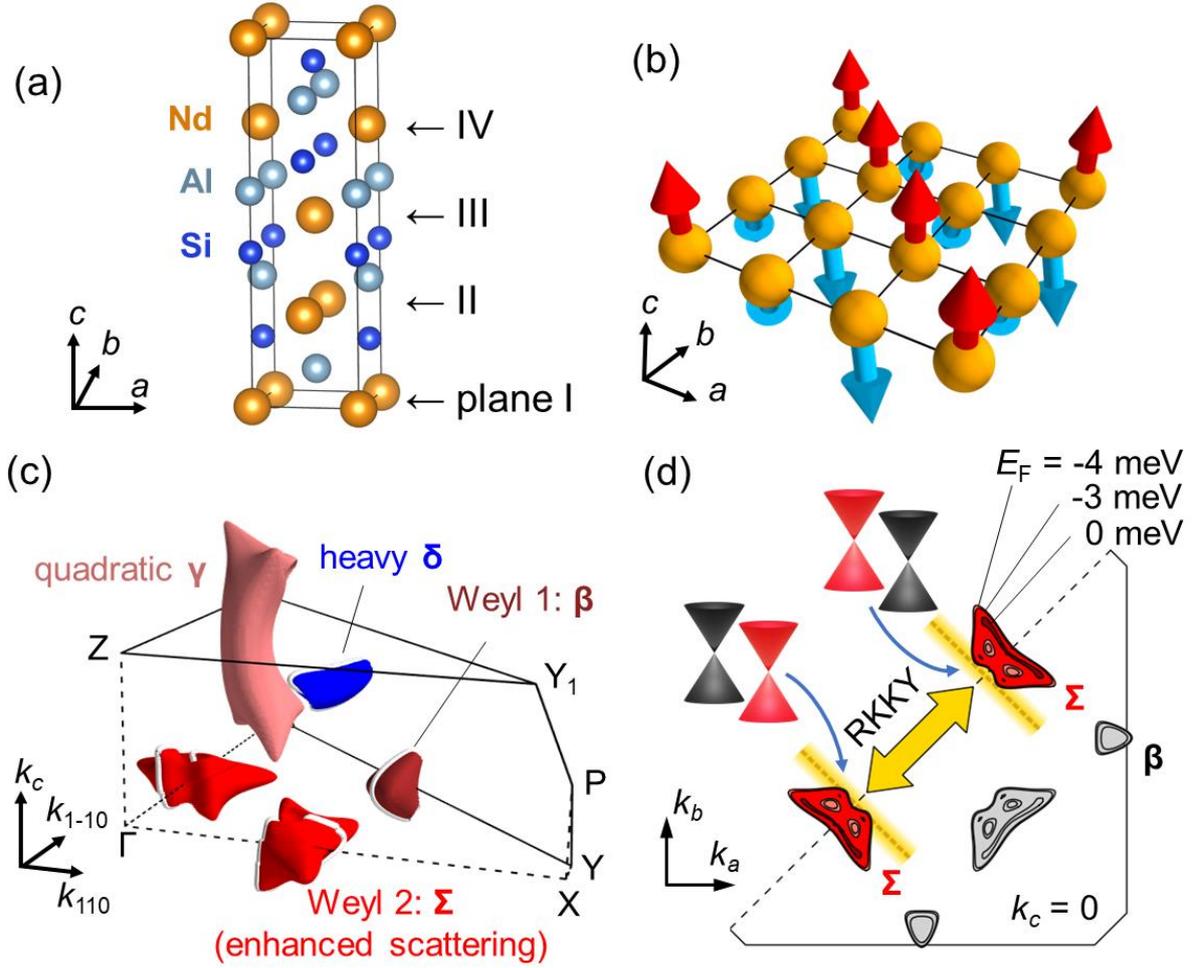

**Figure 1. Fermi surface nesting of Weyl points in NdAlSi.** (a) Crystal structure of NdAlSi, with polar *c*-axis. (b) Spin texture in the *ab*-plane (plane I) with the magnetic ordering vector **Q** ≈ (2/3, 2/3, 0) along the [110] direction, as revealed in Ref. [22]. (c) The electronic structure in NdAlSi; one octant of the Brillouin Zone is shown. There are four types of Fermi surfaces: a high-mobility Weyl β pocket (brown), a Weyl Σ pocket subject to Ruderman-Kittel-Kasuya-Yosida (RKKY) interactions (red), a hole γ pocket with quadratic electronic band dispersion (pink), and an electron δ pocket with heavy electron mass (blue). In the Hall and Nernst effects, only the former three contribute dominantly (below). White solid lines indicate the extremal cross-sections of electron orbits observed experimentally when a magnetic field is applied along the *a*-axis. (d) Two-dimensional cuts of Fermi surfaces at $k_z = 0$, where traces

corresponding to $E_F$ = -4, -3, and 0 meV are shown. Yellow lines and a shaded region indicate the nesting plane with ordering vector **Q**. Illustration: Weyl cones in the electronic band structure of Σ.

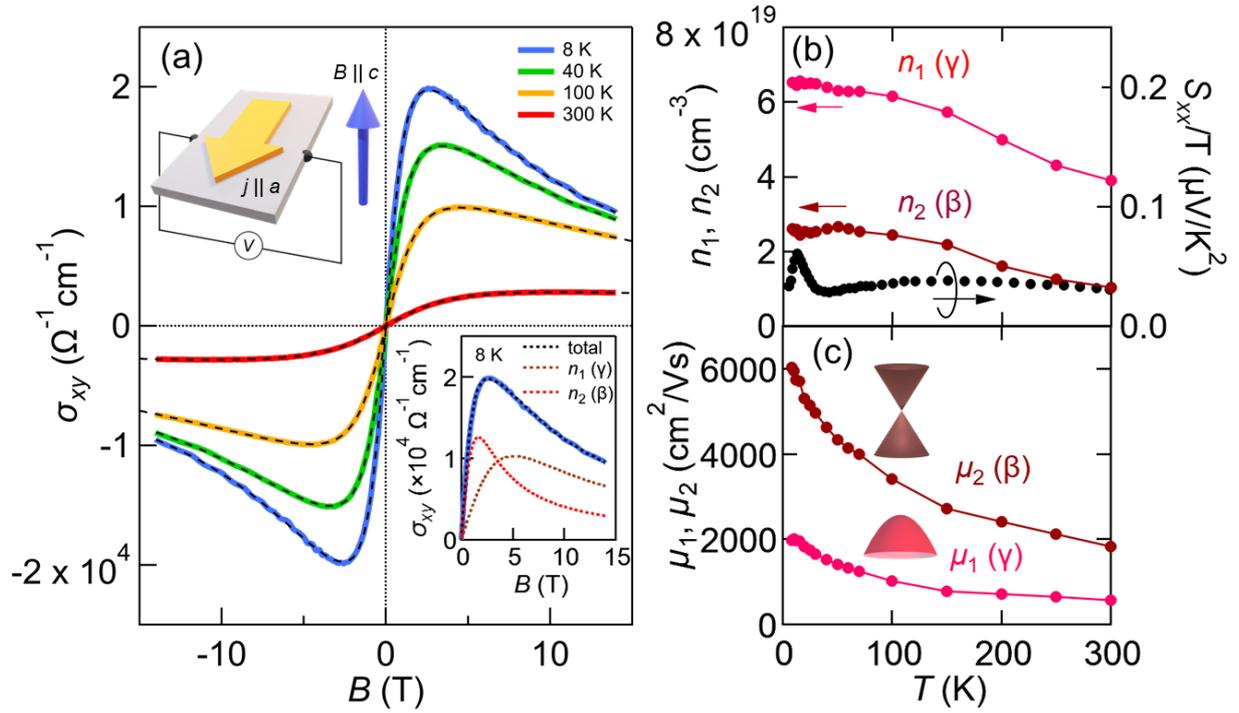

**Figure 2. High mobility electrons from Weyl and trivial Fermi surface pockets in NdAlSi, detected by electrical transport.** (a) Hall conductivity $\sigma_{xy}$ at various temperatures. The black dashed lines indicate fitting to a two-carrier model. Inset (top left): Sample geometry, where magnetic field **B** and electric current **J** are applied along the *c*-axis and *a*-axis, respectively. Inset (bottom) shows the two components of the fit individually. (b) Carrier density of two Fermi surface sheets, estimated by two carrier fitting (left axis). The parameters $\mu_1$, $n_1$ are assigned to a trivial (γ) and $\mu_2$, $n_2$ to a Weyl-type (β) Fermi surface sheet, respectively. Right axis: Temperature dependence of the Seebeck effect divided by temperature, with enhancement above $T_N$ likely due to electron correlations on Σ. (c) Carrier mobility of the trivial γ and Weyl β bands as a function of temperature.

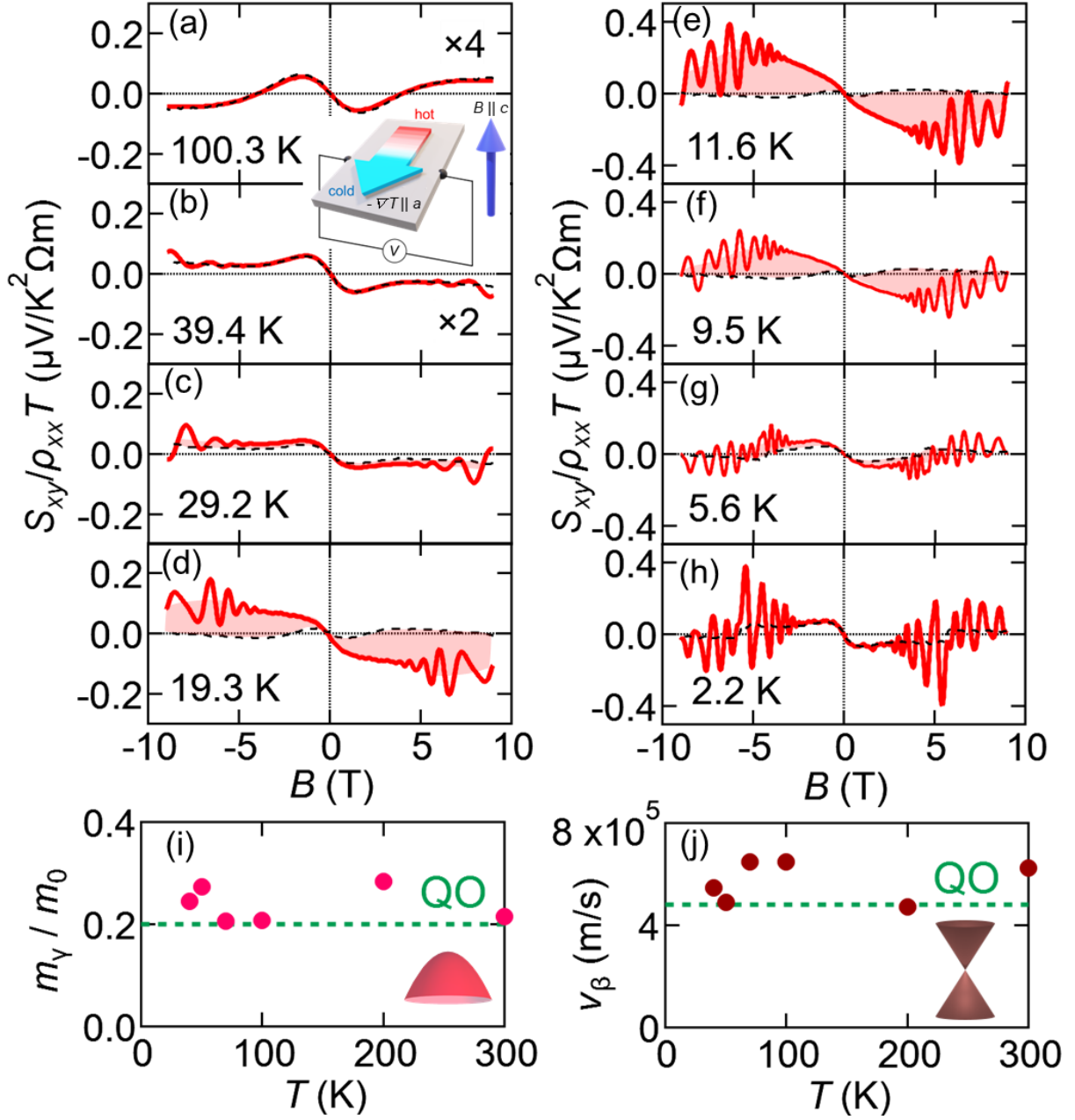

**Figure 3. Two carrier analysis of thermoelectric Nernst effect in NdAlSi.** (a-h) Magnetic field dependence of Nernst effect $S_{xy}$ normalized by resistivity and temperature. Two-carrier decomposition of the Nernst effect is shown by black dotted lines; this represents a fit to Eq. (1) at $T \geq 35$ K and a calculated curve according to Eq. (1), with extrapolated model parameters, at lower temperatures. An excess Nernst signal $\Delta S_{xy}$, beyond Eq. (1), appears at intermediate temperatures $T = 10 - 30$ K, marked by pale red shading between the data and the semiclassical

calculation. Inset: Sample geometry for thermoelectric measurements, where magnetic field and temperature gradient are applied along the crystallographic *c*- and *a*-axes, respectively. (i,j) Temperature dependence of effective mass and Fermi velocity for γ and β pockets with quadratic and linear band dispersions, respectively. Corresponding electronic structure parameters estimated from quantum oscillation (QO) experiments are also shown as green dotted lines in Figs. 3(i) and 3(j).

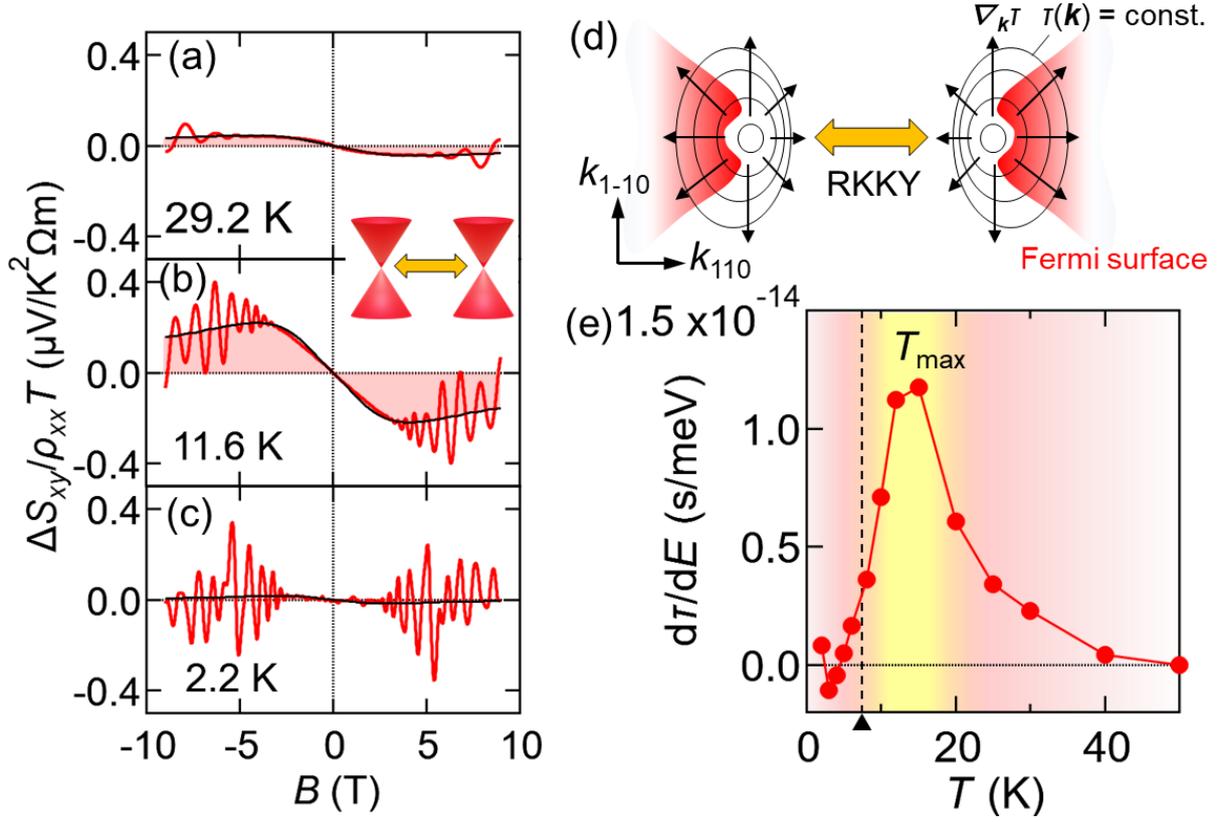

**Figure 4. Enhancement of thermoelectric Nernst effect due electron correlations from a magnetic Fermi surface instability.** (a-c) Magnetic field dependence of excess Nernst signal $\Delta S_{xy}$, normalized by temperature and longitudinal resistivity $\rho_{xx}$, at 30 K, 12 K, and 2 K. A fit to Eq. (2) is indicated by the black line. (d) Illustration: Contour lines of relaxation time $\tau(\mathbf{k})$ in momentum space with a hot-spot of $\tau(\mathbf{k})$, such as can be driven by magnetic ordering with propagation vector $\mathbf{Q} \approx (2/3, 2/3, 0)$ in NdAlSi, c.f. Fig. 1. The arrows indicate a $\mathbf{k}$-space gradient of the relaxation time. For a given electron or hole on the Fermi surface, the effective energy dependence of $\tau$ is a product of this gradient and the inverse Fermi velocity, $d\tau/dE = d\tau/d\mathbf{k} \cdot d\mathbf{k}/dE$. (e) Energy derivative of carrier relaxation time $d\tau/dE$, derived from $\Delta S_{xy}$ and attributed to correlations on the $\Sigma$ pocket. Yellow shading and a vertical dashed line mark the regime of strong fluctuations and the transition to long-range magnetic order at $T_N = 7.4$ K, respectively.

| Material | $\mu$ (cm$^2$/Vs) | $|d\ln(\tau)/dE|$ (eV$^{-1}$) | $|d\ln(E_F)/dE|$ (eV$^{-1}$) | $S_{xx}$ | $S_{xy}$ | References |
|---|---|---|---|---|---|---|
| NdAlSi, β | 6,000 | - | 7 | ○ | ○ | this work |
| NdAlSi, Σ | 3,500 | 33 | 12 | ○ | ○ | this work |
| PbSnSe | 100,000 | - | 20 | ○ | ○ | [17] |
| Co$_{0.999}$Ni$_{0.001}$Sb$_3$ | 200 | 650 | - | ○ | ○ | [12] |
| CeCu$_2$Si$_2$ | 80 | 150 | - | ○ | ○ | [13] |
| CuFeS$_2$ | 60 | 680 | - | ○ | × | [14] |
| Mo$_{0.92}$Nb$_{0.08}$Te$_2$ | 40 | 110 | - | ○ | × | [15] |
| MnGe | ~5 | 10 | - | ○ | × | [16] |

**Table 1. Carrier mobility $\mu$ and energy derivative of the carrier relaxation time $\tau$ for correlated materials with lifetime enhancement of the thermoelectric effect.** The energy derivative of $\mu \sim \tau/E_F$ has two components, from the relaxation time and the Fermi energy, shown in the third and fourth column, respectively [12-17]. Materials with small Fermi surfaces and high carrier mobility, such as PbSnSe, have sizable filling dependence of $\mu$ even in absence of correlations. Among correlated materials with unconventional Nernst responses, NdAlSi shows exceedingly high mobility, which enables us to perform quantitative analysis of $|d\ln(\tau)/dE|$ with resolution in momentum space. Circles (○) and crosses (×) represent if the Seebeck and Nernst effect are measured in the reference or not.

# Supplementary Information for

# Nernst effect of high-mobility Weyl electrons in NdAlSi enhanced by a Fermi surface nesting instability

Rinsuke Yamada *et al.*

*Corresponding author. Email: ryamada@ap.t.u-tokyo.ac.jp or hirschberger@ap.t.u-tokyo.ac.jp

**This PDF file includes:**

Supplementary Text

Figures S1 to S21

Tables S1 and S2

Supplementary References S1-S14

**Supplementary Text**

Raw data of the Nernst effect

Raw data of the Nernst effect divided by temperature ($S_{xy}/T$) at various temperatures is shown in Fig. S1. The Nernst signal $S_{xy}$ for $\bm{B} \parallel c$ follows a non-linear trend versus magnetic field, which enables us to perform a quantitative analysis. The field-dependence of $S_{xy}/T$ has a clear resemblance to $S_{xy}/\rho_{xx}T$ in Figs. 3(a)-(h), including an enhancement around $T_{\max} \sim 15$ K. Thus we exclude the possibility that the anomaly in $S_{xy}/\rho_{xx}T$ originates from an anomaly in $\rho_{xx}$. The raw data of the Nernst effect at 3 T indeed shows an enhancement around 15 K in temperature dependence (see Fig. S1(i)). Also, the Seebeck effect ($S_{xx}/T$) shows a similar enhancement around 15 K (see Figs. 2(b) and S1(j)). Hence, the signal we extracted in Fig. 4(e) is likely robust and not qualitatively dependent on the details of the analysis.

Quantum oscillations and Fermi surface volume

To confirm the consistency of our analysis of the Nernst effect, we extracted band parameters such as carrier density and Fermi velocity from quantum oscillations in the Nernst effect $S_{xy}$ [Fig. S2(a)]. We calculate a fast Fourier transformation (FFT) of the oscillatory component $S_{xy}^{\text{Osci.}}$ after subtracting a smooth polynomial background $S_{xy}^{\text{BG}}$ [Fig. S2(b)]. The spectra show peaks at three frequencies: 51 T, 67 T, and 128 T [Fig. S2(c), Table S1], which are assigned, respectively, to the $\Sigma$, $\beta$, and $\gamma$ pockets in Ref. [S1].

We estimate the carrier density $n$ of each pocket by using the elliptical approximation $n = g_s g_d k_F^a k_F^b k_F^c / 6\pi^2$, where $g_s$ and $k_F^\alpha$ ($\alpha = a, b, c$) are spin degeneracy and Fermi momenta, respectively. The parameter $g_d$, termed orbital degeneracy in the following, counts the number of symmetry-equivalent Fermi surface pockets in the first Brillouin zone. The carrier density of $\beta$ ($\gamma$) at low temperature is estimated to be $1.0 \times 10^{19}$ cm$^{-3}$ ($6.6 \times 10^{19}$ cm$^{-3}$) by considering the $k_F$-anisotropy from the band calculation and by setting $g_s = 1$ ($g_s = 2$) and $g_d = 4$ ($g_d = 4$). The extracted carrier densities from quantum oscillation analysis show a good agreement with those estimated by fitting $\sigma_{xy}$ with a Drude-type semiclassical model (Fig. 2).

From the temperature dependence of Shubnikov-de Haas oscillations in the magnetoresistance, we extract the effective masses for $\Sigma$ and $\beta$ pockets as 0.15 $m_0$ and 0.11 $m_0$, respectively (Fig. S15(a)). This is in good agreement with Ref. [S1], which also determined the effective mass of $\gamma$ to be 0.2 $m_0$. By utilizing the relation $m v_F = \hbar k_F$, the Fermi velocity of $\beta$ is estimated as $4.8 \times 10^5$ m/s. These parameters from quantum oscillations show an excellent agreement with those from our $S_{xy}$ analysis (Figs. 3(i) and 3(j)).

Electron-type carriers in NdAlSi in tunnel diode oscillator (TDO) measurements

We measured the field-dependent change of a tunnel diode oscillator's (TDO) resonant frequency to precisely detect quantum oscillations in pulsed magnetic fields up to 60 Tesla at the Institute for Solid State Physics (ISSP, The University of Tokyo) [S2]. The TDO is based on an $LC$-tank circuit powered by a tunnel diode that is biased in the negative resistance region of its current-voltage characteristic. A coil of about 5 turns and diameter of 1 mm is turned from copper wire. A NdAlSi single crystal is fixed inside the coil by Apiezon-N grease. The coil's axis and the crystallographic $a$-axis of NdAlSi are set perpendicular and in parallel to the magnetic field direction, respectively. We approximate the circuit's resonant frequency as $f \sim 1/2\pi\sqrt{LC}$ with inductance $L$ and capacitance $C$, typically lying in the MHz range. The measurement principle is as follows: The skin depth of a metallic sample is given by $d = \sqrt{\rho/\pi\varepsilon_0 f}$, where $\rho$ and $\varepsilon_0$ are the sample resistivity and the permittivity of the vacuum. Therefore, resistivity variations of the sample result in a change of $LC$-circuit conductance $2r_s\Delta d/R^2$, where $r_s$, $R$, and $\Delta d$ are the sample mean diameter, the coil diameter, and the skin depth variation. This means any change in $f$ is proportional to the variation in sample resistivity, at least to the leading order.

The $LC$-circuit's resonance frequency shows a variation $\Delta f$ as a function of the magnetic field, with pronounced oscillations [Fig. S3(a)]. This quantum oscillation is discerned more clearly in the second derivative of $\Delta f$, shown in Fig. S3(b). We perform separate fast Fourier transformations (FFTs) in the low-$B$ regime (12-25 T) and in the high-$B$ regime (25–60 T), after subtracting a polynomial background from $\Delta f$. The FFT spectra in each regime are shown in Figs. S3(c,d). At low fields, two frequencies 61 T and 104 T are observed, which are assigned to the $\Sigma$ and $\beta$ pockets, respectively (***B*** // *a*). Oscillations with larger frequency (321 T and 125 T) appear at elevated fields, which we denote as $\delta_1$ and $\delta_2$ and which correspond to two different carrier orbits around different projections of symmetry-equivalent, electron-like $\delta$ surfaces [Fig. S3(d), Fig. 1(c), Table S1]. The axis of the oscillation orbit is always perpendicular to the *a*-axis in this experiment. We further analyze the attenuation of the high-field oscillation amplitude against temperature to extract the effective mass of $\delta$ [Fig. S3(e)]. The masses $m(\delta_1)$ and $m(\delta_2)$ are found to be 1.74 $m_0$ and 1.14 $m_0$, respectively, one order of magnitude larger than the reported masses for the hole pockets $\beta$, $\Sigma$, and $\gamma$ [S1]. The high mass of $\delta$ is well consistent with band calculations and in accord with the observation that electron-like carriers, with large effective mass and low carrier mobility, do not contribute significantly to the Nernst and Hall effects.

Change in Hall resistivity across the magnetic transition

We show the raw data of the Hall resistivity at various temperatures in Fig. S7. The Hall resistivity increases with decreasing temperature down to 50 K, however, it becomes nearly temperature independent at low temperatures below 30 K [Figs. S7(a) and S7(b)]. The absence of clear anomaly in temperature evolution of the Hall resistivity at low temperatures suggests that the change in electronic structure at magnetic transition is not significant.

Furthermore, we investigated the possible signature of the anomalous Hall effect by analysis of the field profile of $\rho_{yx}$ at 2 K [Fig. S8]. Although the field dependence of magnetization shows a jump at 0 T due to the flipping of the *c*-component of magnetic moment between two ferrimagnetic states (up-up-down and up-down-down), a clear jump is not observed around 0 T in the Hall resistivity. At the critical field around 6 T, a small step appears in the Hall resistivity. However, the normal Hall effect is much larger than this anomaly due to the small carrier density in the present compound.

Fixing the ratio of mobilities ($\mu_\beta/\mu_\gamma$) and coefficients ($A_\gamma'/A_\beta'$)

Here we explain in detail how we constrained the parameters to obtain reasonable fitting results for the Nernst effect $S_{xy}(B)$. At high temperatures above 40 K, $S_{xy}$ is fitted by the two-carrier model in the semiclassical picture described in Eq. (9) in the main text. Out of four free parameters in Eq. (9), the first constraint is applied by fixing the ratio of mobilities ($\mu_\beta/\mu_\gamma$), considering the fact that the ratio of mobilities from the analysis of Hall conductivity ($\mu_2/\mu_1$) is nearly constant against temperature [see Figs. S9(a) and S9(b)].

The next constraint is the ratio of the coefficients $A_\gamma'/A_\beta'$. We calculated the ratio $A_\gamma'/A_\beta'$ by using the coefficients $A_\gamma'$ and $A_\beta'$ obtained from the analysis of the Nernst effect under the condition of $\mu_\beta/\mu_\gamma = 3.1$ [Figs. S9(c) and S9(d)]. Since the ratio $A_\gamma'/A_\beta'$ shows little temperature dependence from 40 K to 100 K, we fix $A_\gamma'/A_\beta' = 7.0$ for the final analysis of the Nernst effect, to reduce the number of free parameters in our model. We note that $A_\gamma'/A_\beta'$ deviates from 7.0 at 200 K and 300 K; this is likely another local minimum in the fit, caused by the weak field dependence of the Nernst effect due to reduced mobility at very high temperatures. In Figs. S9(e) and S9(f), we show a multi-band fit of the Nernst effect $S_{xy}(B)$ at 300 K, without and with fixing the ratio $A_\gamma'/A_\beta'$. In both cases, the Nernst effect is well explained by the two-carrier model, so we prefer to fix the ratio $A_\gamma'/A_\beta'$ to obtain consistency between 300 K and the lower temperature data.

Extrapolation of the mobility and Nernst amplitude coefficient toward low temperatures

We explain how the two-carrier analysis was extrapolated towards low temperatures. Figure S10 shows the two free parameters obtained from the two-carrier fit of the Nernst effect at high temperatures above 40 K with Eq. (9), fixing the ratio of mobilities and coefficients as discussed above. We fix the value of $A_\gamma'$ to 3.2 when extending the analysis toward low temperatures, assuming that smooth evolution of the mobility and scattering time as a function of temperature and no strong electronic structure changes at the phase transition into the ordered state. The assumption (b) is motivated by the good agreement of the calculated and experimentally observed Nernst effect at $T = 2$ K. Moreover, the thermopower $S_{xx}/T$ in zero magnetic field shows an enhancement around $T_N$ (Fig. 2, akin to the $\Delta S_{xy}$ in Fig. 4), before returning to a comparable value as at high temperature.

Next, we discuss the robustness of the extrapolated parameters for the $T = 2$ K data. We have calculated the sum of square deviation between the data at $T = 2$ K and our model,

$$|\chi|^2 = \sum_{i=0}^{N} \left| \frac{S_{xy}}{\rho_{yx}T}(B_i)_{sim} - \frac{S_{xy}}{\rho_{yx}T}(B_i)_{obs} \right|^2 / N \qquad (S1)$$

where $N$ is a number of data points of the magnetic field [Fig. S11]. Here, we vary the two extrapolated parameters while the ratios $\mu_\beta/\mu_\gamma$, $A_\gamma'/A_\beta'$ remain fixed, as for the high-temperature fits. We indicate the extrapolated value by an orange star and the optimal fit to the low-temperature data by a white circle. Besides the reasonable match of these two, we note that there are no secondary minima in the space of these two parameters.

Fitting of the Nernst signal by the two-carrier model with four free parameters

We show the obtained parameters from the fitting of the Nernst effect by the two-carrier model with four free parameters ($A_\beta$, $A_\gamma$, $\mu_\beta$, and $\mu_\gamma$) in Fig. S12. Naturally, at high temperatures this model gives results that are equivalent to the analysis presented in the main text (the number of adjustable parameters at each temperature has been increased in this fitting with four free parameters). However, at low temperatures below 30 K, sharp anomalies in the parameters $A_1$, $A_2$, and $\mu_2$ indicate inadequacies of this model with four free parameters. As discussed in the main text, this inadequacy of the four-band fit is well resolved by including the $d\tau/dE$ term for $S_{xy}$.

In this analysis and also the analysis in the main text, we neglect the energy dependence of $\tau$ due to scattering processes that have uniform behavior in momentum space. Electron-electron and electron-phonon scattering rates at high temperatures are proportional to the total density of states ($D$), which is more stable (Fig. S13) as compared to the $dn_i/dE$ of each individual band.

Also, for elastic scattering from charge-neutral impurities (limit of strong screening in a good conductor), $\tau$ has weak energy dependence [S3-S6]. This argument can be supported at a quantitative level. The magnitude of $d\ln D/dE$ in our NdAlSi crystal is estimated to be 3 eV$^{-1}$ (Fig. S13), which is one order of magnitude smaller than $d\ln n_i/dE$ (e.g., $d\ln n_\beta/dE \sim 21$ eV$^{-1}$). Also, our estimation of $d\ln \tau/dE$ ($\sim 33$ eV$^{-1}$) of the $\Sigma$ pocket is one order of magnitude larger than $d\ln D/dE$ [see Table 1 in the main text], indicating that the relaxation time contribution is enhanced when the scattering has large a $\boldsymbol{k}$-dependence due to the nesting properties of the Fermi surface.

Analysis of longitudinal conductivity to check for consistency

As a further crosscheck of our two-carrier analysis of the Hall conductivity, we perform a fit to the longitudinal conductivity at various temperatures [Fig. S14(a)]. The experimental data are well described by the two-carrier Drude model as follows:

$$\sigma_{xx} = \frac{n_1 e \mu_1}{1+(\mu_1 B)^2} + \frac{n_2 e \mu_2}{1+(\mu_2 B)^2}. \tag{S2}$$

Here, $n_1$ and $n_2$ ($\mu_1$ and $\mu_2$) are carrier densities (carrier mobilities). The carrier mobilities ($\mu_1$ and $\mu_2$) from $\sigma_{xx}$ show good agreement with those from $\sigma_{xy}$ [Fig. S14(c)]. $n_1$ and $n_2$ at 8 K are estimated to be $8.0\times10^{19}$ cm$^{-3}$ and $4.7\times10^{19}$ cm$^{-3}$ from the fitting of $\sigma_{xx}$, while those from $\sigma_{xy}$ are $6.5\times10^{19}$ cm$^{-3}$ and $2.6\times10^{19}$ cm$^{-3}$, respectively. The observed difference in carrier densities, especially for $n_2$, may originate from a sum contribution of the β and Σ pockets. The effective parameters $n_i$ and $\mu_i$ of the Drude model are expected to capture contributions from several Fermi surface segments, with slightly different carrier mobility. As mentioned in the main text, the Weyl fermions in the $\Sigma$-pocket, may also represent a minority contribution to $n_2$ and $\mu_2$. Such a sum of contributions from β and $\Sigma$ is weighted slightly differently in $\sigma_{xx}$ and $\sigma_{xy}$, possibly explaining the discrepancy in the $n_2$. A second possibility is the experimental uncertainty of sample geometry measurement, which is inevitable for contacts made with Ag paste, which have a width of about a few hundred μm. We assess the typical error bar of 10% due to shape measurement for $\rho_{xx}$ and $\rho_{yx}$; this carries over to $\sigma_{xx}$ and $\sigma_{xy}$, and the relative error can be on the order of 20%.

Mass analysis of quantum oscillations for the investigation of change in electronic structure at the magnetic transition

In order to discuss the possibility that the $dn/dE$ term may explain the enhanced thermoelectric responses in $S_{xx}$ and $S_{xy}$, we perform the mass analysis of quantum oscillations.

The energy derivative of carrier density for quadratic dispersion and for linear band dispersion in three-dimensional materials is described as

$$\frac{dn}{dE} \propto m^* k_F, \qquad \text{and} \qquad \frac{dn}{dE} \propto \frac{k_F^2}{v_F}, \qquad (S3)$$

respectively. A strong change in $dn/dE$ with temperature is therefore directly associated with a change of the effective mass $m^*$, or the Fermi velocity $v_F$. A change of $dn/dE$ can occur and result in a change in thermoelectric responses, even if $n$ itself remains relatively stable over a given temperature window.

Figure S15 shows $T$-dependent SdH oscillations, based on our own analysis (left) and on measurements in a previous report [S1]. The SdH oscillations of the $\beta$, $\Sigma$, and $\gamma$ pockets are all observed below and above $T_N$, and especially in the regime of 10-15 K that is important for our discussion. The smooth, $T$-dependent decay of the oscillation amplitude suggests that any change of the effective masses ($m_\beta$ and $m_\gamma$) is weak between the low-$T$ ferromagnetic regime (suppressed fluctuations) and the higher-$T$ paramagnetic regime (with fluctuations). Therefore, it is unlikely that $dn/dE$ term alone explains the enhanced thermoelectric responses in $S_{xx}$ and $S_{xy}$.

On the need for $d\tau/dE$ to describe the Nernst effect at $T_N < T < 40$ K

We estimate the effect of a reasonable change of the electronic structure on $dn/dE$, and the resulting change of the Nernst effect. We calculate the two-carrier decomposition of the Nernst effect from Eq. (S4), which is equivalent to Eq. (1) of the main text:

$$\frac{S_{xy}}{\rho_{xx}T} = A'_\gamma F(\mu_\gamma, B) + A'_\beta G(\mu_\beta, B). \qquad (S4)$$

The contributions of $\gamma$ and $\beta$ to the Nernst effect at $T = 11.6$ K are shown in Fig. S16 as green and purple dotted lines, respectively (simulation without the effect of $n_2$); the blue solid line represents their sum. Considering the possible change in the carrier density $n_2$ of the Weyl $\beta$ pocket by about 15 % [Fig. 2], we here assume a change of $A_\beta'$ by 20% and calculate the resulting change of $S_{xy}(B)$, to exaggerate the effect. When adjusting $A_\beta'$ by 20 %, we obtain the red dashed line, only marginally different from the previous estimate and far away from explaining the large enhancement of $S_{xy}$ just above $T_N$ (black curve), which is the focus of our work. More generally, we note that the observed $S_{xy}$ is considerably much larger than each individual component to Eq. (S4); the signal enhancement cannot be explained even if $A_\beta'$ and $A_\gamma'$ show some more drastic variation around $T_{max}$. We conclude that our analysis of $S_{xy}$ is not qualitatively changed by the reduction in carrier density $n_2$ or by moderate changes in $A_\beta'$, $A_\gamma'$.

Extraction of relaxation time of each Fermi surface

We focus on three relatively light, hole-like carrier pockets of NdAlSi: trivial $\gamma$ and Weyl $\beta$ & $\Sigma$ pockets, where only the latter ($\Sigma$) is close to the nesting condition in RKKY theory. In Fig. S17, we present the carrier relaxation times $\tau$ extracted from the mobilities, together with the relaxation time $\tau_\Sigma$ of the $\Sigma$ pocket as obtained from the fit to $\Delta S_{xy}(B)$ according to the last term in Eq. (8). A clear anomaly at $T_{max}$ ~15 K appears for $\Sigma$. We caution that the quantitative value of $\tau_\Sigma$ of the $\Sigma$-pocket is more sensitive to details of the analysis than $\tau_\beta$ or $\tau_\gamma$.

To support Fig. S17(a), we also estimated the quantum scattering time $\tau_Q$ of the Weyl $\beta$ pocket from an analysis of the dominant Shubnikov-de Haas (SdH) oscillation frequency. A fit to the Lifshitz-Kosevich formula is shown in Fig. S17(c) [S7]. At 8 K we find $\tau_Q = 1.2 \times 10^{-13}$ s [Fig. S17(c)], reproducing $\tau_\beta$ from transport experiments within a factor of 3. Note that quantum lifetimes are almost always reduced as compared to transport carrier relaxation times, being equally sensitive to both small- and large angle scattering [S8].

The values of $\tau_\beta$, $\tau_\gamma$, and $\tau_\Sigma$ are scattered within a factor of 2, consistent with moderate or weak correlations in NdAlSi. Even for $\Sigma$, the change of $\tau_\Sigma$ due to magnetic fluctuations is a mild effect on the order of 15% – which, however, significantly affects the thermoelectric response, as $d\tau_\Sigma/dE$ is more sensitive to the correlation effect than $\tau_\Sigma$ itself. While $\tau_\beta$ and $\tau_\gamma$ monotonically increase toward low temperature, $\tau_\Sigma$ shows a dip around $T_{max}$ ~ 15 K, in good agreement with our scenario. As the RKKY interaction couples electronic states separated by the ordering vector $\boldsymbol{Q}$ in momentum space, it is natural that the Weyl $\Sigma$ pocket is affected more strongly than $\beta$ and $\gamma$ [Figs. 1(c) and 1(d) of the main text].

More quantitatively, we can cross-check our estimate for $d\tau_\Sigma/dE$ shown in Fig. S17(b) [same as Fig. 4(e) in the main text] using the temperature-induced change of $\tau_\Sigma$ in Fig. S17(a). The maximal reduction $max(|\Delta\tau_\Sigma|)$ at $T_{max}$=12 K is estimated to be approximately $5 \times 10^{-14}$ s as compared to a smooth background of the same shape as $\tau_\beta(T)$. In our analysis, we found that the maximum $d\tau_\Sigma/dE$ appears at 15 K, and assumed that this is when the minimum of $\tau(E)$ is just at the HWHM of the Fermi-Dirac distribution centered around $E_F$ (Fig. S6). Therefore, we calculate 2 × HWHM × $max(d\tau_\Sigma/dE)$ to estimate the maximum change of $\tau_\Sigma$ at the perfect nesting condition: 2×3 meV×(1.2×10$^{-14}$ s/meV) = 7.2×10$^{-14}$ s, which is close to two times the size of $max(|\Delta\tau_\Sigma|)$.

Fitting of the field dependence of the excess Nernst signal

Here we explain in detail how we attribute the excess Nernst signal ($\Delta S_{xy}$) to the $\Sigma$ pocket close to the nesting condition. The field dependence of the additional component in $S_{xy}$ is fitted by

$$\frac{\Delta S_{xy}}{\rho_{xx}T} = \sum_i B_i \left( \frac{\pm 2\mu_i B - \tan\theta_H (1-(\mu_i B)^2)}{(1+(\mu_i B)^2)^2} \right), \qquad (S5)$$

which contains the contributions from the energy dependence of the relaxation time ($B_i = -g_s g_d e k_B^2 k_i^3/18m_i \cdot d\tau_i/dE$). The upper and lower sign notation describes hole- and electron-type carriers, respectively. By fitting $\Delta S_{xy}$ assuming the contributions of hole-type and electron-type carriers (see Figs. 4(a-c) and S18(a-c)), we estimate the relaxation time $\tau_i = \mu_i m_i/e$ for each case, and plot them against temperature in Fig. S18(d). When we assume that $\Delta S_{xy}$ originates from electron-type $\delta$ pocket, the relaxation time is estimated to be $1.3 \times 10^{-12}$ s. This value is not physically reasonable because it is more than three times larger than that for the Weyl $\beta$ pocket with a small effective mass (~4×10$^{-13}$ s). Therefore, we choose hole-type carrier contributions as the solution for the fitting of $\Delta S_{xy}$.

Furthermore, the temperature dependence of the relaxation time of hole contributions $\tau_{hole}$ (black) is suppressed around 15 K, although $\tau_\beta$ (pink) and $\tau_\gamma$ (light brown) monotonically increase toward the lowest temperature (see Fig. S18(d)). It is true that the estimation of $\tau_{hole}$ may contain a large error bar, the difference in the temperature dependence suggests that $\Delta S_{xy}$ comes from the pocket other than $\beta$ or $\gamma$ pocket, i.e. from the $\Sigma$ pocket.

Observation of a lifetime-driven Nernst effect in magnetic Weyl semimetal GdPtBi
    $R$AlSi and $R$PtBi ($R$ = rare earth) both break inversion and time-reversal symmetries in their long-range ordered magnetic state. Both compounds show strong magnetic fluctuations above $T_N$~7-9 K, and both have elevated carrier mobilities (~1,000-5,000 cm$^2$/Vs). A particular advantage of GdPtBi is its simple electronic structure, namely its low carrier concentration, its quadratic band touching at the $\Gamma$-point of the first Brillouin zone, and its Fermi energy tunability with small amounts of Au/Pt doping.

We report the Nernst effect of two samples of GdPtBi with different band filling, controlled by Au-doping: Stoichiometric GdPtBi (Sample S01) and Au-doped GdPt$_{0.985}$Au$_{0.015}$Bi (Sample S02), where the magnetic field was applied along the (111) direction and the heat current was // (110). We show the low-field slope of the Nernst effect divided by temperature in Figs. S19(e) and S19(f).

At base temperature, GdPtBi with quadratic band touching is dominated by a single carrier type, implying that the normal Nernst effect should vanish due to Sondheimer cancellation. At high temperatures, thermally activated carriers appear, resulting in a strong temperature dependence of the Nernst effect (~d$n$/d$E$). This differs from semimetallic NdAlSi, where the carrier density is much larger and the normal Nernst effect (~d$n$/d$E$) shows weak temperature dependence; thermally activated carriers play a lesser role in NdAlSi, and both electron- and hole-type carriers are present at all temperatures [no Sondheimer cancellation, see Fig. S19(d)]. There are key similarities between the two materials, however: akin to NdAlSi, the Nernst signal of GdPtBi shows an anomaly close to the Curie-Weiss temperature $T_{CW}$, around 30 K.

In order to separate the contributions of d$n$/d$E$ and d$\tau$/d$E$, we fitted the temperature dependence of the Nernst effect of GdPtBi, focusing on the low field regime ($\mu B \ll 1$). Here, the Nernst effect for the two-carrier model can be derived from Eq. (9) in the main text as

$$\frac{S_{xy}}{\rho_{xx} B T} = \frac{(3\pi^2)^{\frac{1}{3}} k_B^2}{3\hbar^2 e^2} \frac{m_1}{\rho_{xx}^2}\left(1+\frac{1}{b}\right) \frac{\frac{1}{b} n_1^{\frac{1}{3}} n_2 + n_2^{\frac{1}{3}} n_1}{\left(n_1+\frac{1}{b} n_2\right)^3} + \frac{\Delta S_{xy}}{\rho_{xx} B T}, \qquad (S6)$$

where $n_1$, $n_2$, $m_1$, $m_2$, $\mu_1$, and $\mu_2$, are the carrier density, effective mass, and mobility of carrier 1 (holes) and 2 (electrons), respectively. The first term in Eq. (S6) originates from the energy derivative of carrier density d$n$/d$E$, while the second term represents the additional contribution including the relaxation time effect. We introduce a constant "$b$" as the ratio of carrier masses ($b = m_2/m_1$).

To fit the temperature dependence, we eliminate some of these parameters, making reasonable assumptions: First, given the quadratic band touching and obeying Luttinger's rule, we enforce constant $n_1$-$n_2$ at all temperatures and set the electron density $n_2$ to be activated with an activation barrier $E_a$. Second, we use the same relaxation time for both carriers. Third, we set the mass ratio to $b = 0.25$, based on band structure calculations and prior experimental work [S9]. This means the electrons are significantly much lighter than the holes.

Blue lines in Figs. S19(g)-(i) show fits following the contribution from $dn/dE$ at high temperatures with three adjustable parameters: For hole-type S01 (slightly electron-type S02), we obtain $E_a$ = -15 meV (+7 meV) defining the activation-type behavior at low $T$ as measured from the band-touching point, and $m_1$ = 0.7~1.3 $m_e$ from the overall signal amplitude.

A deviation from the model for $dn/dE$ appears at low temperatures, see red highlights in Figs. S19(g)-(i). For $B$ // (111), as here, the heat current is moving in planes of triangular lattice arrangement of Gd ions in the half-Heusler structure, precluding a contribution from scalar spin chirality fluctuations. Therefore, as in NdAlSi, we attribute the additional Nernst signal that onsets close to the Curie-Weiss temperature $T_{CW}$ to magnetic fluctuations and the relaxation time effect. We conclude that such thermoelectric phenomena should appear ubiquitously in correlated topological semimetals, and beyond.

Details on the Nernst effect measurements

The Hall effect and Nernst effect data are anti-symmetrized Hall effect and Nernst effect data with respect to the magnetic field, to eliminate the contribution of the Seebeck effect due to the misalignment of the voltage terminals (see Figs. S20(a) and (b)). Furthermore, we confirmed that the voltage drops of the thermocouples, thermopower, and Nernst terminals show a linear dependence against the heater power ($P = RI^2$, with heater resistance $R$ and current to the heater $I$) as shown in Fig. S20(c), which suggests that the thermal gradient is assessed to be homogeneous within the measured region of our sample.

Reproducibility of the enhanced Nernst effect

We measured two samples to confirm the reproducibility of our results on the enhanced Nernst effect. The Nernst effect divided by temperature ($S_{xy}/T$) for the 2[nd] sample (sample 2) is shown in Fig. S21(a). With decreasing temperature from 52 K, $S_{xy}/T$ increases down to 14.4 K, accompanied by clear quantum oscillations. $S_{xy}/T$ decreases at further lower temperatures and the overall field dependence of sample 2 is similar to that of the 1[st] sample (sample 1), which is discussed in the main text. Both samples show the enhanced Nernst signal around 15 K (see Fig. S1(i) and S21(b)). We note here that the absolute value of $S_{xy}/T$ differs between the two samples (~30%). The difference in absolute signal amplitude likely originates from uncertainties in the measurement of the sample geometry, i.e. in the measurement of the distance between voltage terminals. We also confirmed that the temperature dependence of the Seebeck coefficient and thermal conductivity of sample 2 reproduce the results of sample 1 (Fig. S21(c) and (d)).

Comparison of quantum oscillation frequencies between experiment and density functional theory

Table S1 summarizes the cross-sectional areas of Fermi surfaces from the experiment and DFT calculations. The DFT calculations well reproduce the frequencies observed in the experiment, without a need to significantly shift the energy from the value determined by DFT. Specifically, the characteristic frequencies of $\Sigma$, $\beta$, and $\gamma$ surfaces show good agreement when setting the Fermi energy to 0 meV, while $\delta$ requires a shift to $E_F$= -0.7 meV for a reasonable agreement. Please note that the calculated $\delta$-pocket has a narrow neck at $E_F = 0$, and its characteristic oscillation frequency changes rapidly even with a small shift in $E_F$ (Fig. S3, inset). As compared to a previous study, our work provides a more thorough characterization of oscillation frequencies with the magnetic field along the $a$-axis [S1].

Calculation of the Seebeck effect from band parameters

As for the Seebeck effect $S_{xx}$, compensation between the electron and hole pockets occurs according to Eq. (S2). In the Drude approximation, at zero magnetic field, the Seebeck effect $S_{xx}$ is expressed as

$$S_{xx} \equiv A \frac{\partial \ln \sigma_{xx}}{\partial \varepsilon} = Ae^2 \frac{\partial}{\partial \varepsilon} \ln \sum_i \frac{n_i}{m_i}, \qquad (S7)$$

where $A=-\pi^2 k_B^2/3e$, by ignoring the relaxation time contribution. Here, we assume that the scattering time (at high temperature) is identical for all Fermi surfaces, and independent of momentum. By considering carriers with quadratic dispersion ($E = \hbar^2 k_F^2/2m^*$, $n = g_s g_d k_F^3/6\pi^2$) or linear dispersion ($E = \hbar v_F k_F$, $n = g_s g_d k_F^3/6\pi^2$), we derive

$$S_{xx} = -\frac{\pi^2}{3} \frac{k_B^2}{e} \frac{\sum_{i:\text{Quad}} \left[\pm \frac{1}{2\pi^2 \hbar^2} g_s^i g_d^i k_F^i\right] + \sum_{i:\text{Line}} \left[\pm \frac{1}{3\pi^2 \hbar^2} g_s^i g_d^i k_F^i\right]}{\sum_{i:\text{Quad}} \left[\frac{1}{6\pi^2 \hbar} g_s^i g_d^i {k_F^i}^2 v_F^i\right] + \sum_{i:\text{Line}} \left[\frac{1}{6\pi^2 \hbar} g_s^i g_d^i {k_F^i}^2 v_F^i\right]}, \qquad (S8)$$

where $g_s$, $g_d$, $k_F$, and $v_F$ are spin degeneracy, orbital degeneracy, Fermi momentum, and Fermi velocity, respectively. The upper and lower signs (±) denote hole- and electron-type carriers, respectively.

Firstly, we calculate $S_{xx}/T$ to be 0.38 μV/K$^2$ from Eq. (S8) assuming parallel conduction of the $\gamma$, $\beta$, $\Sigma$ pockets with hole type carriers, but without taking into account the $\delta$ electron pocket. We use $k_F$ and $v_F$ of these three pockets from the analysis of quantum oscillations, while $g_s$ and $g_d$ are set to the values used in the Nernst analysis. The calculated value of $S_{xx}$ is one order of the magnitude larger than the experimental observation (Fig. 2(b)); as compared to the longitudinal conductivity, carrier compensation leads to a highly sensitive behavior of the thermopower on band parameters in this multi-band system. A reasonable agreement between the experiment and calculation is obtained by assuming a moderately small spin splitting and reduced spin degeneracy of the $\delta$ electron pocket even in zero magnetic field, due to spin-orbit interactions and the noncentrosymmetric crystal structure of NdAlSi [Table S2].

Estimation of the magnitude of the life-time contribution to the thermoelectric effects

We estimate the size of the relaxation time contribution to the thermoelectric effects in Table 1 as follows. In the case of NdAlSi, $|d\ln(\tau)/dE|$ is calculated by using the extracted $d\tau/dE$ from the analysis of $S_{xy}$ and the relaxation time $\tau$. $|d\ln(\tau)/dE|$ of GdPtBi and GdPt$_{0.985}$Au$_{0.015}$Bi are estimated to be 15 eV$^{-1}$ and 54 eV$^{-1}$, respectively, which are on the same magnitude as that of NdAlSi. $|d\ln(\tau)/dE|$ of other compounds in Table 1 are estimated by assuming that the observed Seebeck effect originates only from the relaxation time contribution, i.e. $S_{xx} = -\frac{\pi^2}{3}\frac{k_B^2}{e}T\frac{\partial \ln \tau}{\partial \varepsilon}$.

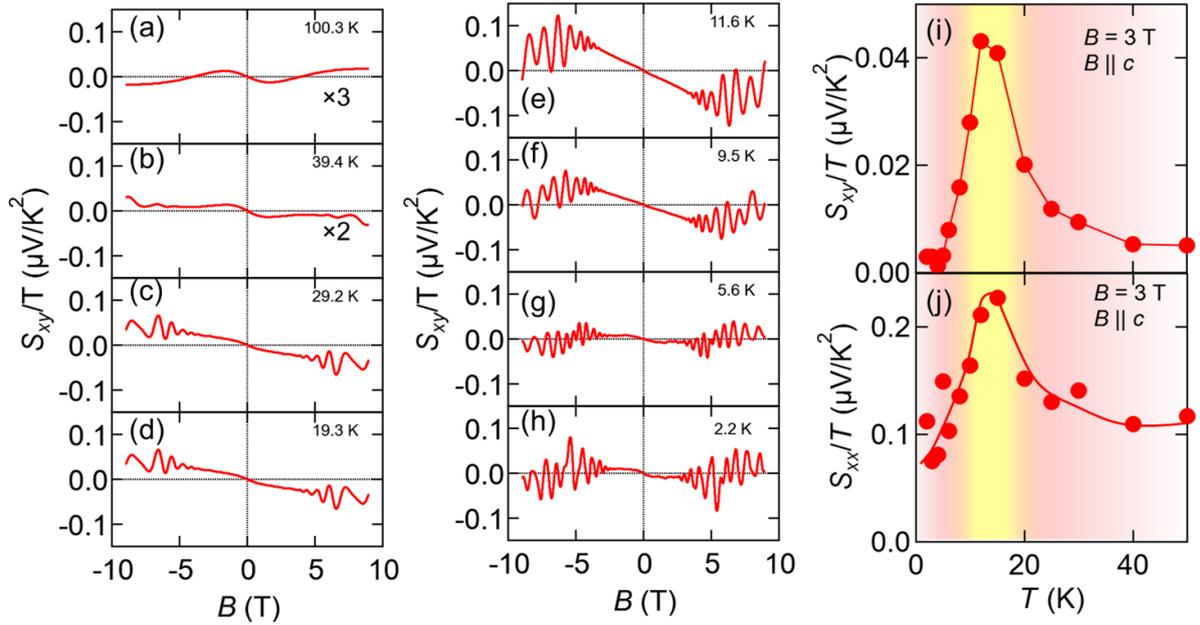

**Fig. S1. Raw data of the Nernst effect of NdAlSi.** (a)-(h) The Nernst effect divided by temperature ($S_{xy}/T$) from 100 K to 2 K. Nernst signal $S_{xy}$ for $B \parallel c$ shows non-linear behavior as a function of the magnetic field, which enables us to perform a quantitative analysis. The overall field dependence resembles the data shown in Figs. 3(a)-3(h). (i, j) Temperature dependence of $S_{xy}/T$ and $S_{xx}/T$ at 3 T. The red curve in (b) is the guide to the eye. An enhancement around 15 K is observed in $S_{xy}/T$ and $S_{xx}/T$ at 3 T, which is similar to the enhanced Seebeck effect at 0 T around 15 K (see Fig. 2(b)).

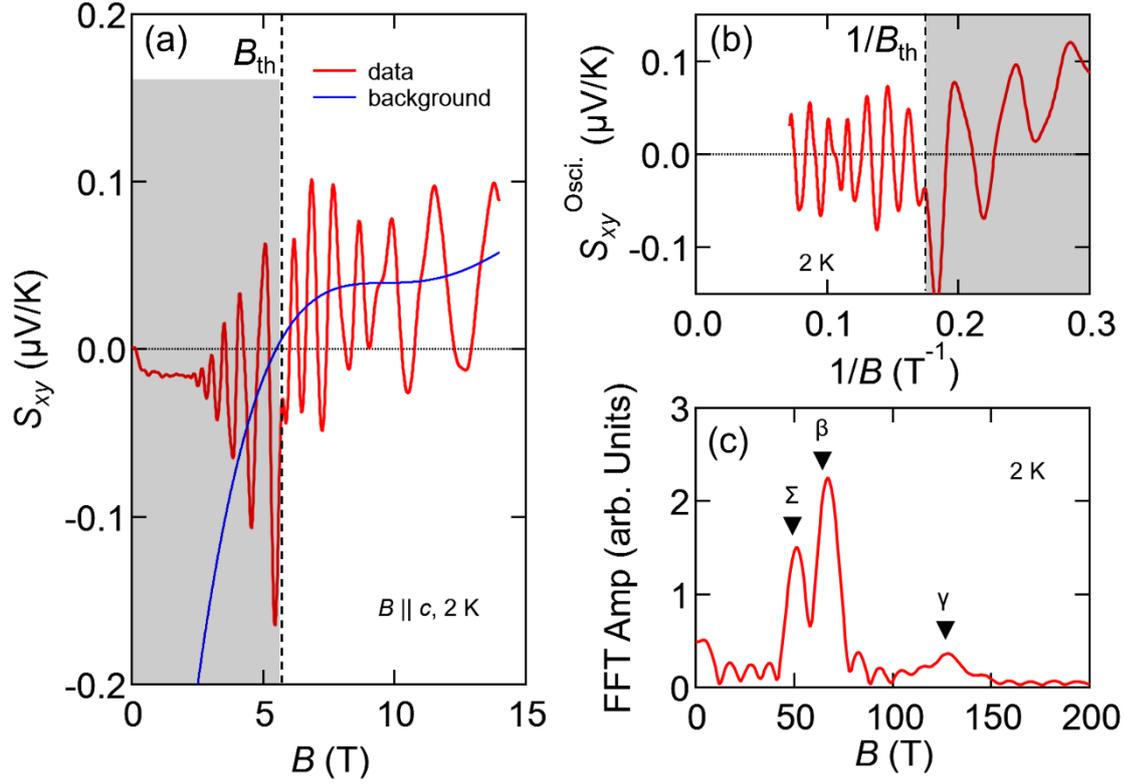

**Fig. S2. Quantum oscillations in the Nernst effect. a,** Nernst effect $S_{xy}$ at 2 K for ***B*** // *c*. We focus on quantum oscillations in the field-aligned ferromagnetic phase above the critical field $B_{th}$, where Nd spins are coaligned. The blue line represents the background from a non-oscillatory component $S_{xy}^{BG}$. **b,** Oscillatory component $S_{xy}^{Osci.}$ as a function of $1/B$, obtained by subtracting $S_{xy}^{BG}$ from $S_{xy}$. **c,** A fast Fourier transformed spectrum of $S_{xy}^{Osci.}$. Three frequencies of 51, 67, and 128 T are observed, which were assigned, respectively, to the $\Sigma$, $\beta$, and $\gamma$ pockets first identified in Ref. [S1]. The grey shaded area indicates the ferrimagnetic regime.

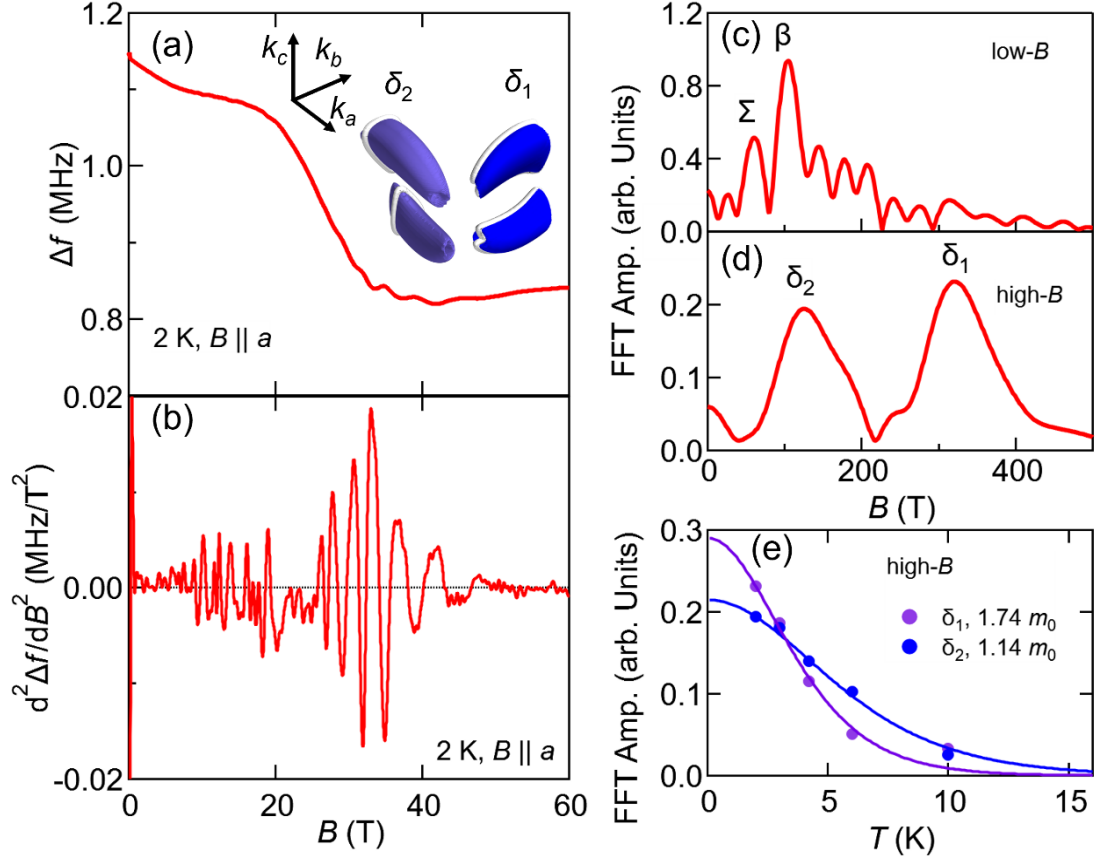

**Fig. S3. Quantum oscillations of NdAlSi in tunnel diode oscillator (TDO) measurements in pulsed magnetic fields. a**, Isotherm of the resonant frequency $\Delta f$ of a tank circuit whose inductance is sensitive to changes in the sample's penetration depth for MHz electromagnetic fields (tunnel diode oscillator, TDO measurement). **b**, Second derivative of $\Delta f$ with respect to magnetic field, demonstrating clear quantum oscillations. **c,d**, Fourier spectra of the oscillatory part of $\Delta f$ in the low field (10-20 T) and high field (25 to 60 T) regions. **e**, Temperature dependence of the Fourier amplitude, and estimation of the effective mass for two orbits on the heavy-electron $\delta$ pocket in the high-$B$ regime. The inset to **a** shows the two cyclotron orbits of the $\delta$ pocket for $B \parallel a$.

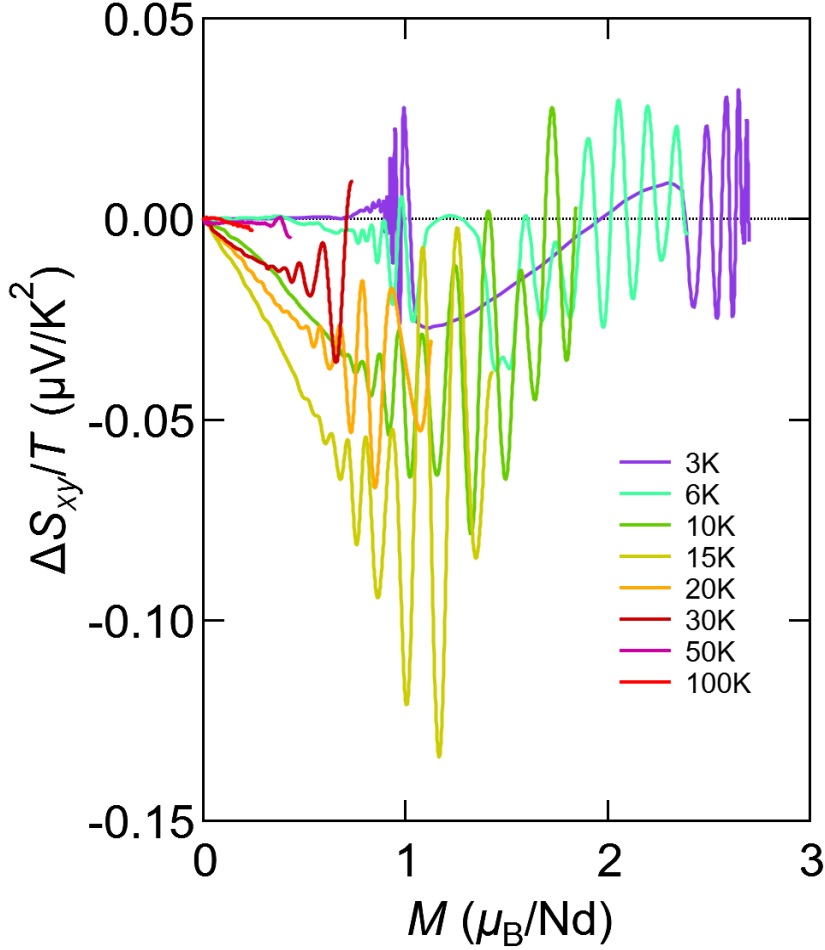

**Fig. S4. The excess Nernst signal ΔS_xy in NdAlSi does not scale with the bulk magnetization.** Excess Nernst effect divided by temperature $\Delta S_{xy}/T$ plotted against magnetization $M$. Clear scaling behaviour onto a universal curve is not observed, in contrast to materials with a thermoelectric Nernst effect from spin-chirality driven Berry phases [S10-S13]. Even while the magnetization is still far from saturation and magnetic fluctuations are expected to persist, the present $\Delta S_{xy}$ nearly vanishes. Instead, the signal is well described by Eq. (2). Note that, when focusing on individual square net layers of Nd ions in NdAlSi, scalar spin chirality of magnetic fluctuations is expected to cancel between neighbouring squares.

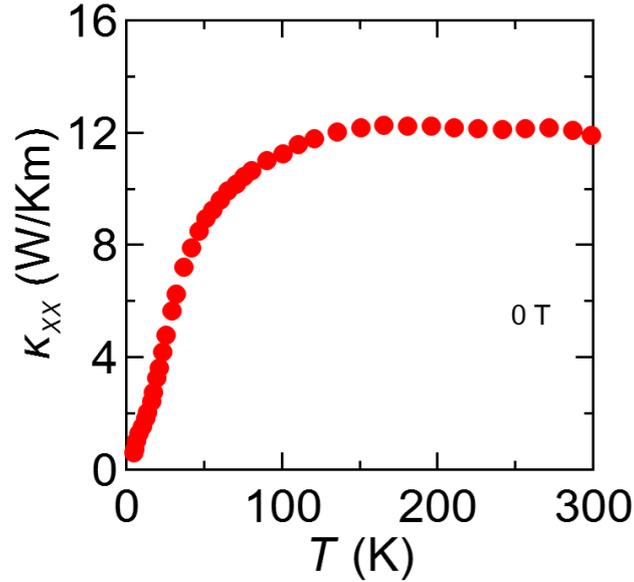

**Fig. S5. Thermal conductivity $\kappa_{xx}$ and absence of phonon drag in NdAlSi.** Smoothly decreasing towards the lowest temperatures below 50 K, the longitudinal thermal conductivity does not show a pronounced peak around 15 K – unlike the Seebeck effect in Fig. 2(b). Combined with temperature-independent behaviour of $S_{xx}/T$ above 35 K, this data is consistent with the near-absence of phonon drag contributions to the thermoelectric properties, where lattice vibrations couple to charge carriers and enhance the thermoelectric Seebeck or Nernst effects [S14].

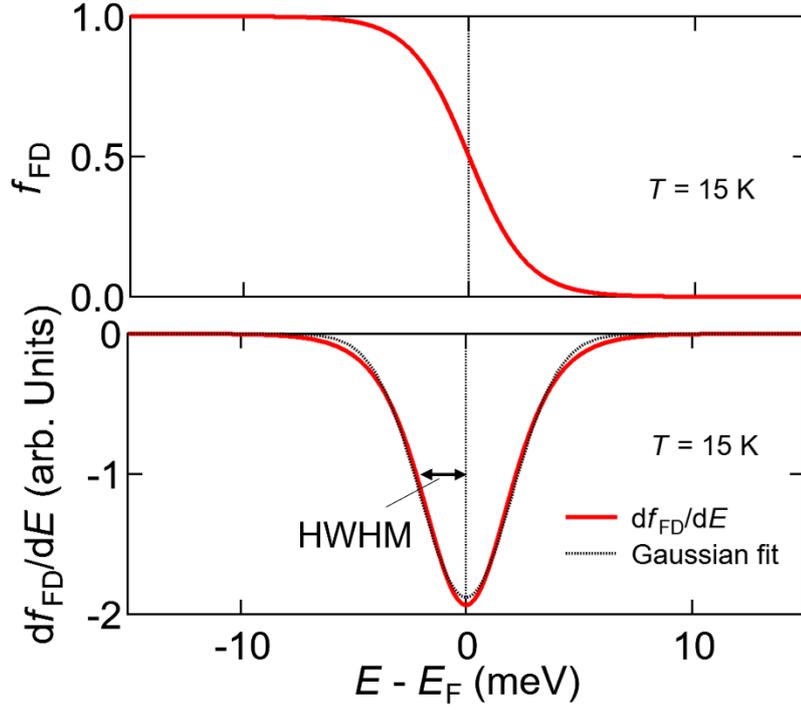

**Fig. S6. Fermi-Dirac distribution function ($f_{FD}$) and its half width at half maximum (HWHM).** From the energy derivative of the Fermi-Dirac distribution function $f_{FD} = 1/(\exp((E-E_F)/k_BT)+1)$ at 15 K, the HWHM is about 3 meV.

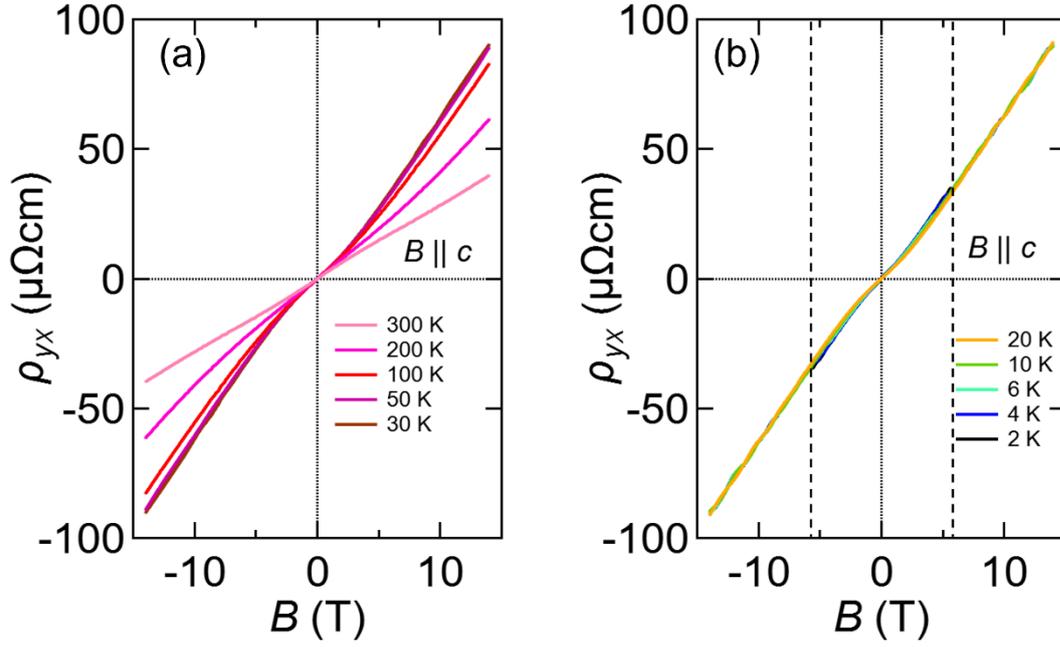

**Fig. S7 Hall resistivity at various temperatures.** Hall resistivity $\rho_{yx}(B)$ of NdAlSi (a) at high temperatures and (b) at low temperatures. The change of the Hall resistivity across the magnetic transition temperature is very hard to quantify, and even up to 15 K, where we observed the enhanced Nernst signal, we could not observe a clear change. In panel (b), the critical field at 2 K is denoted by dashed lines.

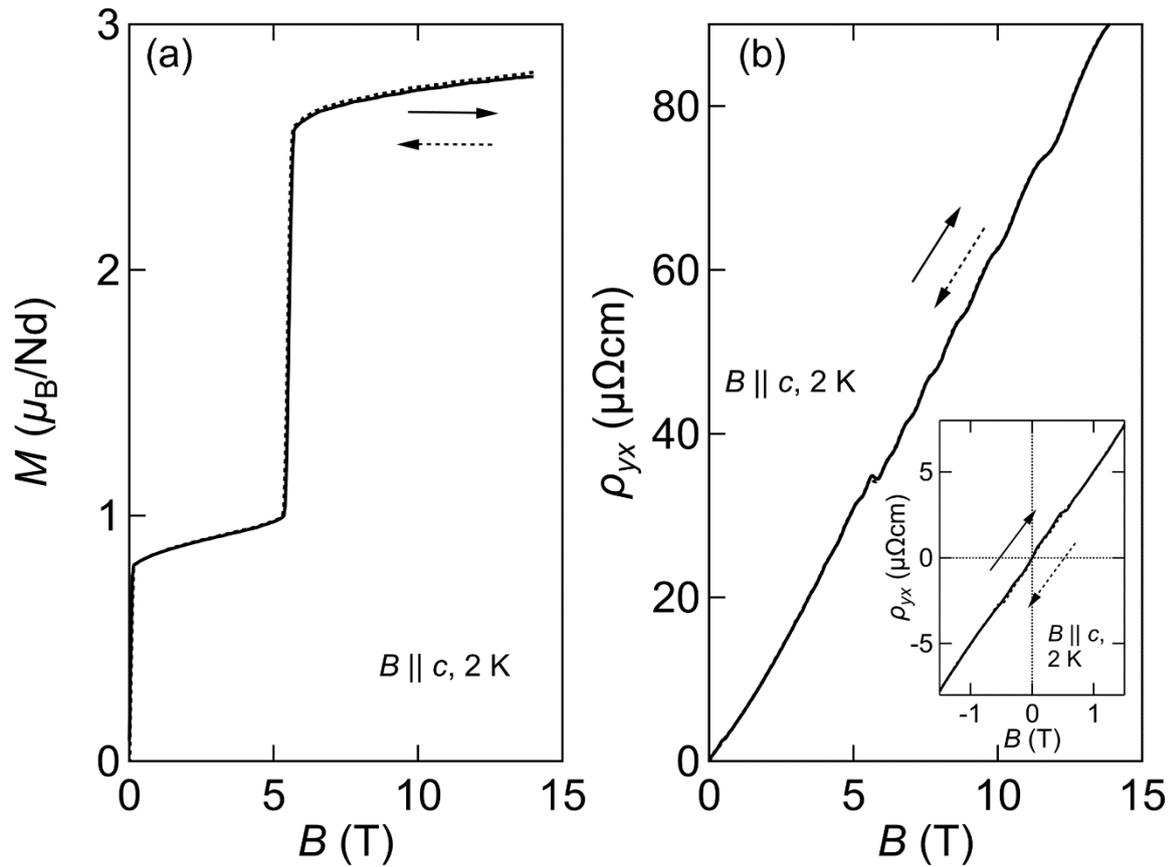

**Fig. S8 Hall effect of NdAlSi at low temperature.** (a) Magnetization and (b) Hall resistivity of NdAlSi at 2 K for $B \parallel c$. The inset shows the magnified view around 0 T without clear signatures of anomalous Hall effect. The solid and dashed lines indicate increasing and decreasing magnetic field (they mostly overlap precisely). The signal is mostly dominated by the normal (field-linear) Hall effect.

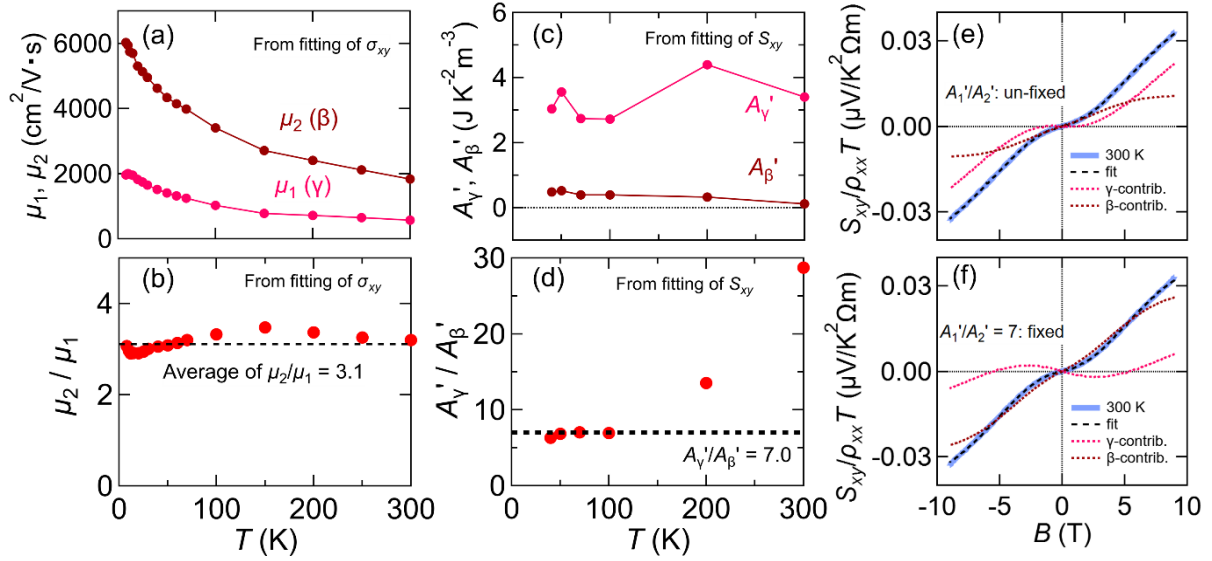

**Fig. S9. How the number of free parameters was constrained in our model.** (a) Temperature dependence of mobilities of the $\gamma$-pocket ($\mu_1$) and $\beta$-pocket ($\mu_2$) extracted from two carrier fitting of the Hall conductivity, without constraint. (b) The ratio of the mobilities ($\mu_2/\mu_1$) is nearly constant against temperature. (c) The coefficients in Eq. (9) of the contribution from the $\gamma$-pocket ($A_\gamma'$) and $\beta$-pocket ($A_\beta'$). (d) The ratio of the two coefficients ($A_\gamma'/A_\beta'$) and the dashed line, which corresponds to the assumption in the (constrained) fit of the Nernst effect. (e, f) Nernst effect at 300 K is fitted by the two-carrier model (e) without constraints on the coefficients $A_1'$ and $A_2'$, and (f) by fixing the ratio $A_1'/A_2'$. The field dependence of the Nernst coefficient can be well described in both cases, however, we chose the results of (f), because small number of free parameters are required.

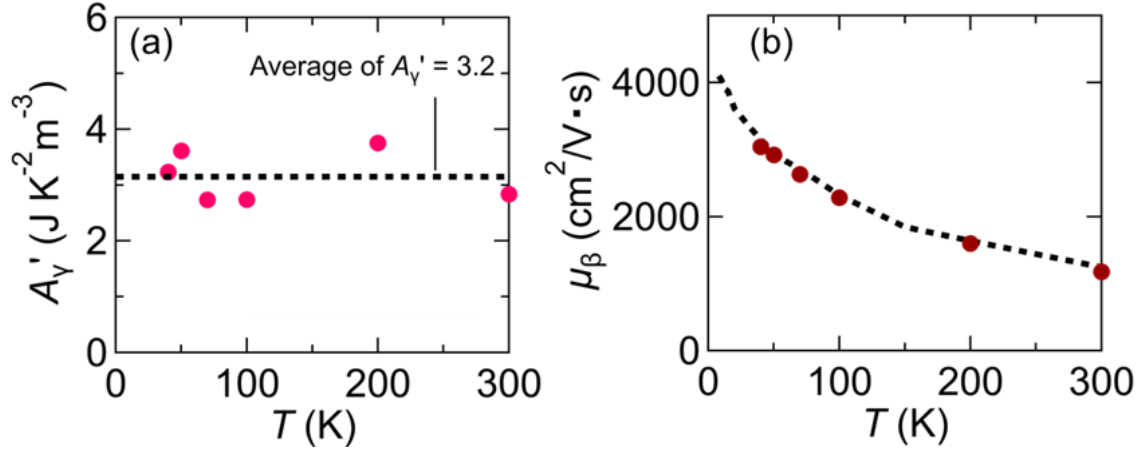

**Fig. S10. Extrapolation of the analysis toward low temperatures.** (a) The coefficient $A_\gamma'$ obtained from the fitting of the Nernst effect. We fixed the value of $A_\gamma'$ as 3.2 when extending the analysis toward low temperatures, as discussed in the text. (b) Mobility $\mu_\beta$ extracted from the fitting of the Nernst effect (symbols) and $\mu_2$ obtained from the analysis of $\sigma_{xy}$ (dotted line, multiplied by a factor of 0.68). To extrapolate $\mu_\beta$ in the Nernst effect to low temperature, we followed this dashed line.

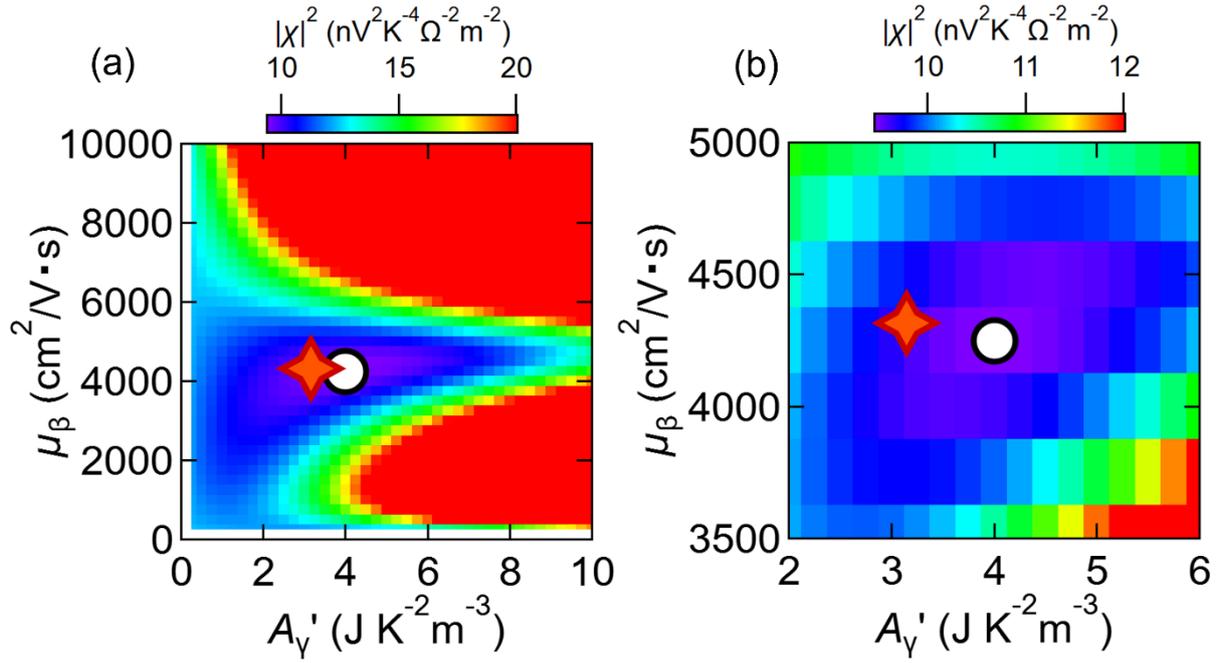

**Fig. S11. Robustness of the extrapolated parameters.** The square deviation of the simulated curve from the experimental data at 2 K ($|\chi|^2$) plotted as a function of $\mu_\beta$ and $A_\gamma'$ under the condition that the ratio of mobilities and prefactors are fixed as $\mu_\beta/\mu_\gamma = 3.1$ and $A_\gamma'/A_\beta' = 7.0$. (b) A magnified view of Fig. S11(a) with the expanded color scale. The orange symbol and white circle indicate the parameters we chose and those show a local minimum, respectively.

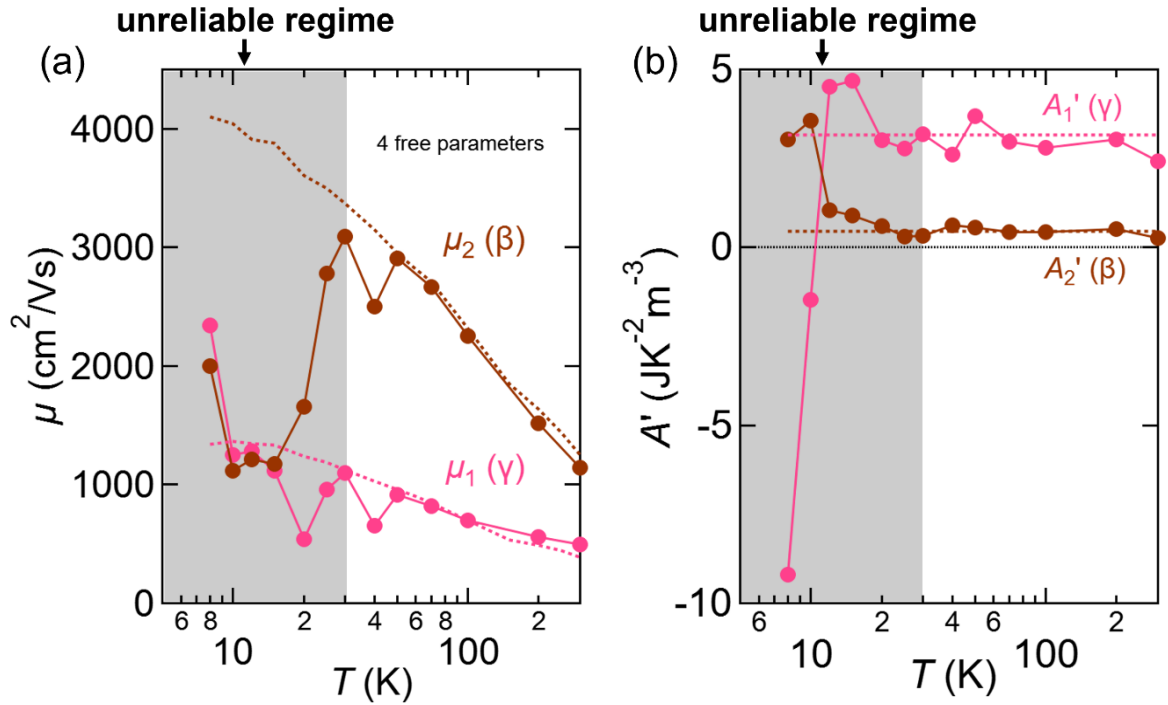

**Fig. S12. Fitting of the Nernst effect by the two-carrier model with four free parameters and without d$\tau$/d$E$ term.** Extracted (a) mobility $\mu$ and (b) coefficients $A'$ from the two-carrier fitting. The calculated mobilities show an anomalous decrease below 30 K, indicating that the two-carrier model is unreliable in this low-temperature regime. Dotted lines are adapted from Fig. S10, for the model presented in the main text.

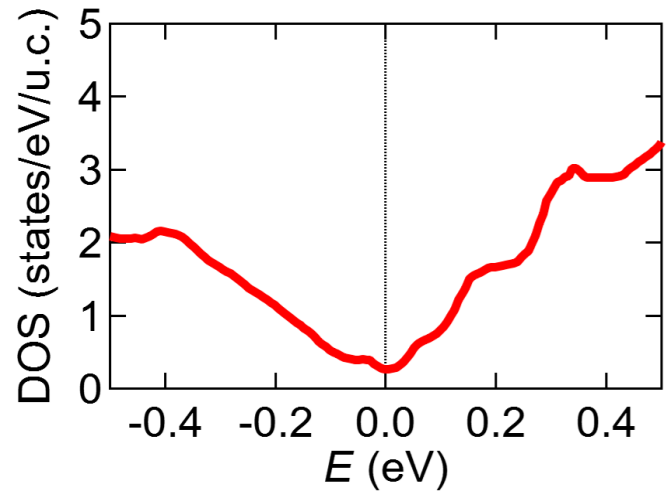

**Fig. S13. Density of states of NdAlSi from DFT calculations.** The total density of states of NdAlSi in the ferromagnetic state, with a local minimum around $E = 0$ eV.

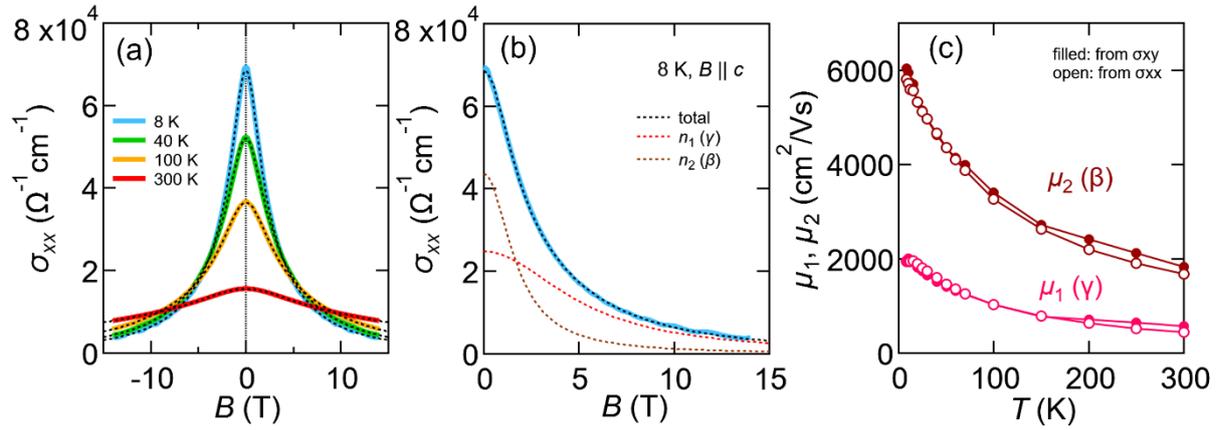

**Fig. S14. Two-carrier Drude fitting of the longitudinal conductivity $\sigma_{xx}$.** (a) Longitudinal electrical conductivity $\sigma_{xx}$ at various temperatures. The black dashed lines indicate fitting to a two-carrier Drude model. (b) Two components of the fit, depicted individually. (c) Carrier mobility of the trivial $\gamma$ and Weyl $\beta$ bands as a function of temperature. Open (filled) symbols denote mobilities derived from the longitudinal electrical conductivity (from the Hall conductivity), respectively.

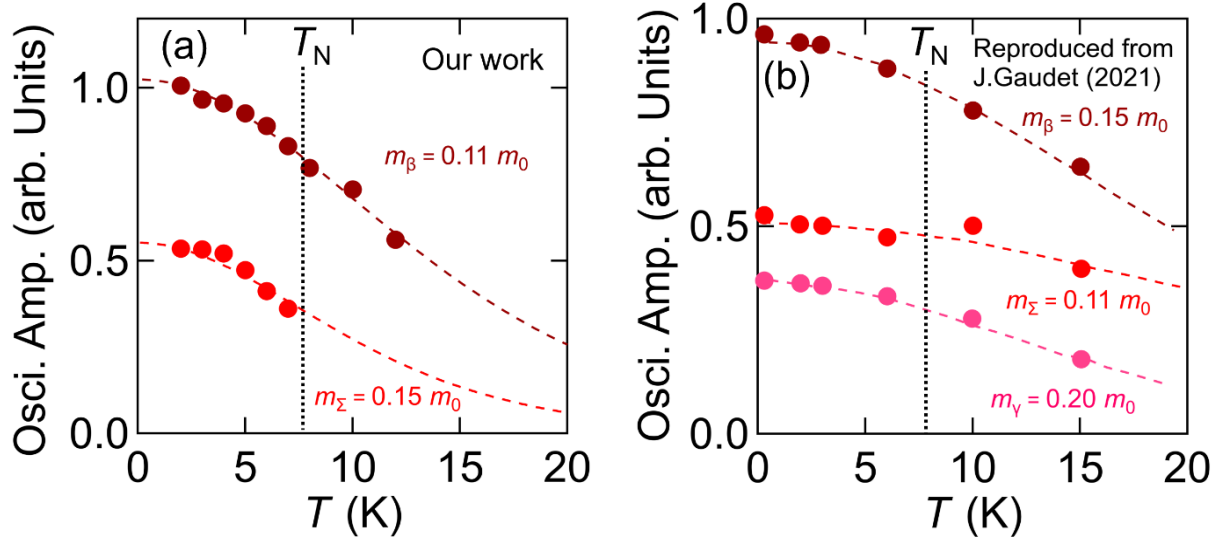

**Fig. S15 Change in electronic structure across the magnetic transition.** Temperature dependence of the FFT amplitude of SdH oscillations (a) in magnetoresistance measured up to 14 T in our work and (b) in magnetoresistance up to 35 T adapted from Ref. S1. The temperature dependence of each pocket can be fitted with a single effective mass across the Néel temperature, indicating that the change of electronic structure is not significant between the field-aligned ferromagnetic and the high-temperature, thermally fluctuating states.

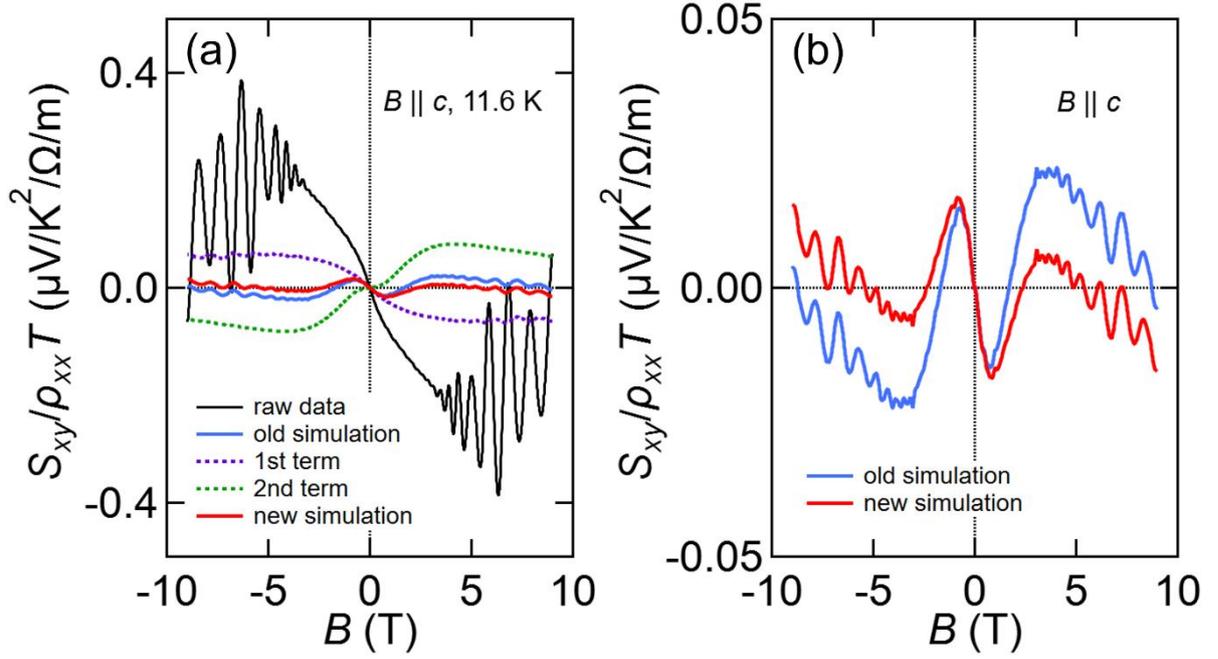

**Fig. S16. Modeling the Nernst effect of NdAlSi above $T_N$ by Drude theory**. (a) Magnetic field dependence of the Nernst effect $S_{xy}$ normalized by resistivity and temperature at $T = 11.6$ K, close to $T_{max}$ (black curve). The blue line represents a two-carrier decomposition of the Nernst effect calculated according to Eq. (9). The contributions from trivial $\gamma$ and Weyl-type $\beta$ pockets are shown as purple (1st) and green (2nd) dotted lines. The red line is the revised two-carrier calculation of $S_{xy}$, where a 20% change of the parameter $A_\beta$ for the Weyl-type $\beta$ pocket is considered. (b) Magnified view of (a), with expanded vertical axis.

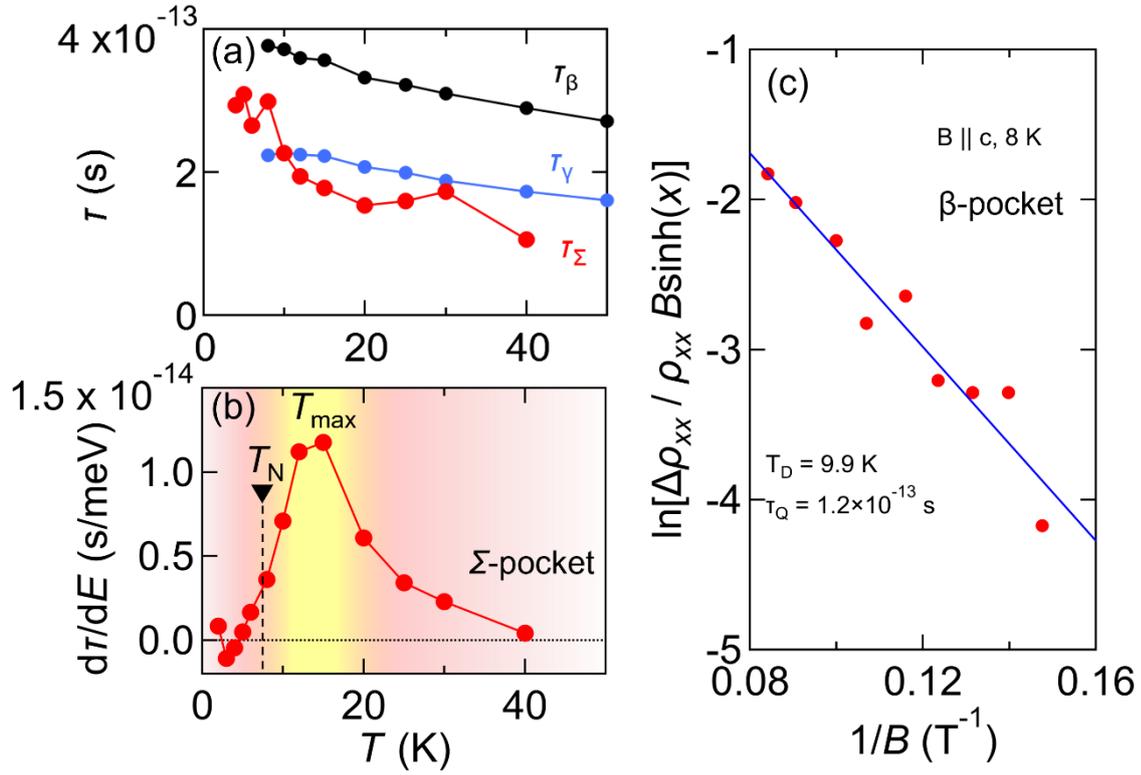

**Fig. S17. Comparison of relaxation time and energy derivative of relaxation time in NdAlSi.** (a) The blue, black, and red points correspond to the relaxation times of the trivial $\gamma$ pocket, the Weyl $\beta$ pocket, and the Weyl $\Sigma$ pocket, respectively. The former two are calculated from the Hall conductivity, and the latter is calculated from a fit to $\Delta S_{xy}$ according to Eq. (8) in the main text. The $\Sigma$-pocket is close to the nesting condition in NdAlSi and shows a dip of $\tau$ around $T_{max} \sim 15$ K. (b) Energy derivative of carrier relaxation time $d\tau_\Sigma/dE$, derived from $\Delta S_{xy}$ and attributed to correlations of electrons on the $\Sigma$ Fermi surface. (c) (b) Cross-check of relaxation time estimate $\tau_Q$ using Dingle plot for Shubnikov-de Haas (SdH) oscillation of the Weyl $\beta$ pocket, showing $\ln[\Delta\rho_{xx}/\rho_{xx}B \sinh(x)]$ versus $1/B$ at 8 K together with a linear fit (blue line). Here, $\Delta\rho_{xx}$ represents the oscillatory amplitude, $x = 2\pi^2 k_B T/\hbar\omega_c$, and $T_D$ is the Dingle temperature.

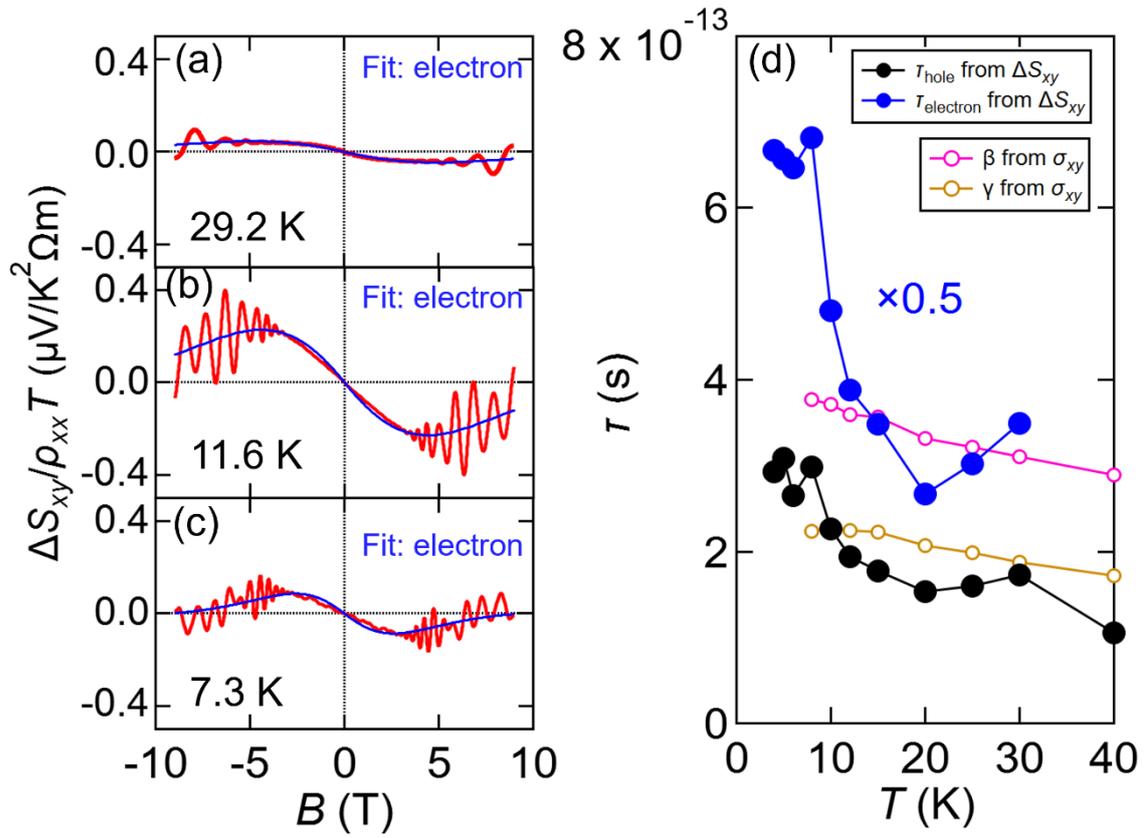

**Fig. S18. Analysis of the field dependence of the excess Nernst signal ($\Delta S_{xy}$).** (a-c) Fitting of $\Delta S_{xy}$ assuming the contribution of electron-type carriers at several temperatures. (d) Extracted scattering time plotted against temperature from the analysis of $\Delta S_{xy}$ and the Hall conductivity.

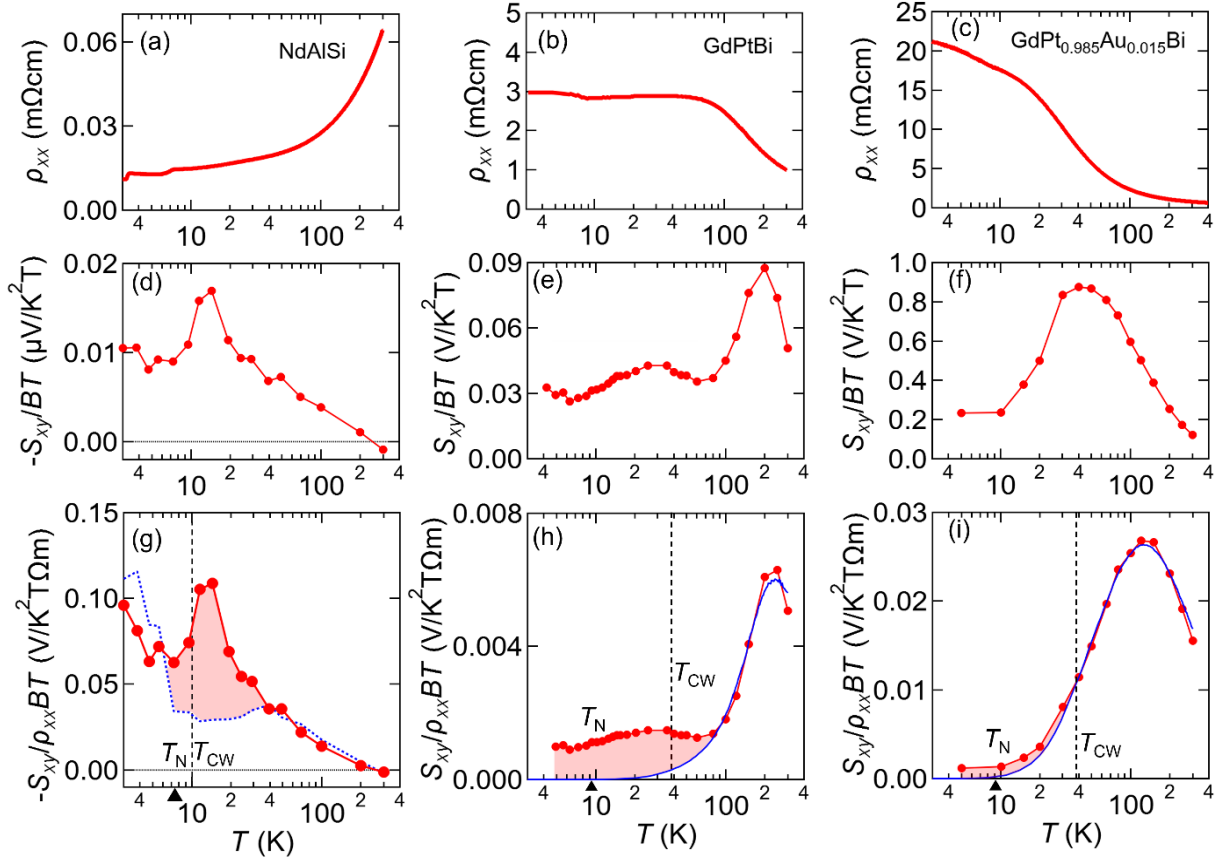

Fig. S19. **Low-field Nernst effect originating from the relaxation time contribution in magnetic topological semimetals NdAlSi and GdPtBi**. (a-c) $\rho$-$T$ curves of semi-metallic NdAlSi, semi-metallic GdPtBi, and semi-conducting GdPt$_{0.985}$Au$_{0.015}$Bi. (d-f) Temperature dependence of the low-field slope of the Nernst effect divided by temperature for three samples, with $\boldsymbol{B} \parallel [111]$ for GdPtBi and $\boldsymbol{B}$ // [001] for NdAlSi. (g-i) Temperature dependence of the low-field slope of $S_{xy}$, divided by resistivity and temperature. The blue dotted lines indicate the contribution from the energy derivative of carrier density d$n$/d$E$. The d$n$/d$E$ of GdPtBi and GdPt$_{0.985}$Au$_{0.015}$Bi are calculated by considering thermally activated carriers around the band-touching point (see text), while that of NdAlSi is estimated by taking the low-field slope of the two-carrier fitting at 0 T (following Fig. 3 of the main text). The red shaded area is attributed to the relaxation time contribution of the Nernst effect ($\Delta S_{xy}$), which is enhanced just above the Néel temperature ($T_N$) denoted by a black triangle. The Curie-Weiss temperature ($T_{CW}$) is shown by a black dashed line. Note that the low-$T$ increase in panel (g) is due to a rapid decrease of resistivity in the low-temperature ordered state, and that fluctuations persist down to lower temperatures in GdPtBi.

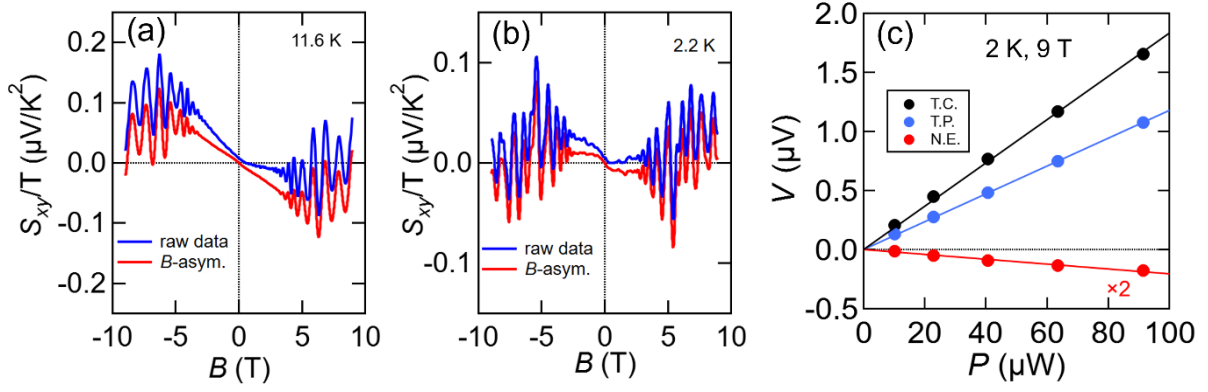

**Fig. S20. Details on the Nernst effect measurements.** (a, b) Field dependence of the Nernst effect divided by temperature ($S_{xy}/T$) at 11.6 K and 2.2 K, respectively. The blue and red curves show the raw data and the data anti-symmetrized to the magnetic field. (c) Heater power dependence of the voltage drop of the thermocouples (T.C.), thermopower (T.P.), and Nernst terminals (N.E.) at 2 K and 9 T. The linear fits are shown to demonstrate the linearity to the heater power. We used the heater power of 40 μW for the measurement at 2 K.

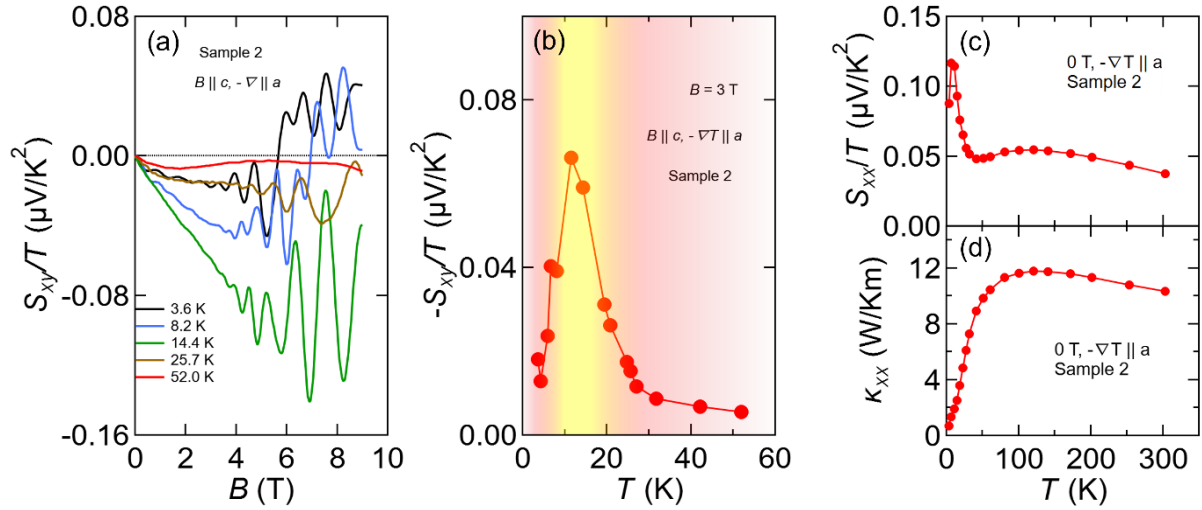

**Fig. S21. Reproducibility of the enhanced Nernst effect.** (a) Magnetic field dependence of the Nernst effect divided by temperature ($S_{xy}/T$) at various temperatures for sample 2. (b) Temperature dependence of $S_{xy}/T$ at 3 T for the sample 2. (c, d) Temperature dependence of the Seebeck effect divided by temperature and thermal conductivity for sample 2.

**Table S1. Comparison of quantum oscillation frequencies between experiment and density functional theory.** Characteristic frequencies of $\Sigma$, $\beta$, and $\gamma$ pockets show good agreement when setting the Fermi energy to 0 meV, while $\delta$ requires a shift to $E_F$= -0.7 meV for a reasonable agreement. Note that $\delta$ has a narrow neck at $E_F = 0$, and its characteristic oscillation frequency changes rapidly even with a small shift in $E_F$ (Fig. S3, inset).

| Fermi pocket | $\Sigma$ ($B \parallel c$) | $\beta$ ($B \parallel c$) | $\gamma$ ($B \parallel c$) | $\delta_1$ ($B \parallel a$) | $\delta_2$ ($B \parallel a$) |
|---|---|---|---|---|---|
| $F^{QO}$ (T) | 51 | 67 | 128 | 321 | 125 |
| $F^{DFT,\ 0\ meV}$ (T) | 49 | 72 | 138 | 802 | 270 |
| $F^{DFT,\ -0.7\ meV}$ (T) | 109 | 83 | 158 | 321 | 105 |

**Table S2. The estimation of thermopower at 0 T.** The Seebeck effect divided temperature $S_{xx}/T$ simulated by assuming parallel conduction of the $\gamma$, $\beta$, $\Sigma$ pockets with hole type carriers without $\delta$ electron pocket and with $\delta$ electron pocket of spin degeneracy $g_s$ = 1, 1.5, and 2, respectively. The experimental value at 0 T and 50 K is also shown. Good agreement between the experiment and calculation requires the small splitting of the $\delta$ electron pocket.

|  | $S_{xx}/T$ (μV/K²) |
| --- | --- |
| Experiment (0 T, 50 K) | 0.034 |
| without $\delta$-pocket | 0.380 |
| $g_s$ = 1 for $\delta$-pocket | 0.133 |
| $g_s$ = 1.5 for $\delta$-pocket | 0.023 |
| $g_s$ = 2 for $\delta$-pocket | -0.079 |

asymmetry in the energy dependence of carrier relaxation time," Chem. Mater. **32**, 2639-2646 (2020).